\documentclass[aps,prd,reprint,superscriptaddress,nofootinbib]{revtex4-2}
\usepackage[all]{xy}
\usepackage{amsmath,amsthm,amssymb}
\usepackage[dvips]{graphicx}
\usepackage{comment}
\usepackage{array}
\usepackage{bm}
\allowdisplaybreaks[1]

\usepackage{xcolor}
\usepackage{hyperref}
\usepackage{tcolorbox}
\usepackage{mathrsfs}
\usepackage{float}
\usepackage[caption=false]{subfig}

\newcommand{\El}{{\sf El}}

\newcommand{\chiinf}{\chi_{\infty}}
\newcommand{\rmin}{r_{\rm min}}
\newcommand{\rcut}{r_{\rm cut}}
\newcommand{\rmax}{r_{\rm max}}
\newcommand{\rsplit}{r_{\rm split}}

\newcommand{\vf}{\varphi}
\newcommand{\lmax}{\ell_{\rm max}}

\newcommand{\lm}{\ell m}
\newcommand{\lw}{\ell\omega}
\newcommand{\lmw}{\ell m \omega}
\newcommand{\wbar}{\tilde{\omega}}

\newcommand{\SP}[1]{\left(#1\right)}

\newcommand{\SB}[1]{\left[#1\right]}

\begin{document}

\title{Frequency-domain approach to self-force in hyperbolic scattering}

\def\Soton{Mathematical Sciences, University of Southampton, 
Southampton SO17 1BJ, United Kingdom}

\author{Christopher Whittall}
\affiliation{\Soton}

\author{Leor Barack}
\affiliation{\Soton}

\date{\today}
\begin{abstract}
We develop a frequency-domain method for calculating the self-force acting on a scalar charge on a fixed scattering geodesic in Schwarzschild spacetime. Existing frequency-domain methods, which are tailored for bound orbits, are inadequate here for several reasons. One must account for the continuous spectrum in the scattering problem, deal with slowly-convergent radial integrals that are hard to evaluate numerically, and confront the inapplicability of the standard self-force method of ``extended homogeneous solutions'', which only works for compactly supported sources.  We tackle each of these issues in turn, and then present a full numerical implementation, in which we calculate the self-force correction to the scatter angle due to scalar-field back-reaction. We perform a range of internal validation tests, as well as ones based on comparison with existing time-domain results. We discuss the merits and remaining limitations of our method, and outline directions for future work. 
\end{abstract}

\maketitle

\section{Introduction}\label{sec:intro}
Calculations of the scatter angle in hyperbolic black-hole encounters have been of recent cross-disciplinary interest. Post-Minkowskian (PM) calculations of the scatter angle may be used to calibrate effective-one-body (EOB) models of binary black-hole mergers \cite{ Damour:2016gwp, Damour:2017zjx, Damour2020, Bini:2020rzn}, providing waveforms that are suitable for gravitational-wave data analysis for operating and upcoming detector experiments. Studies of gravitational scattering are further motivated by the discovery, using effective field theory techniques \cite{ 
PhysRevD.73.104029}, of so-called “boundary-to-bound” relations connecting scatter observables to bound-orbit observables \cite{ Damour:2017zjx,Neill_2013,Cheung:2018wkq,Kalin:2019rwq,Kalin_2020, ChoKalin2021}. 
The problem has drawn significant attention from outside the traditional gravitational physics community, leading to the introduction of new techniques and rapid developments in the PM theory of two-body dynamics in recent years \cite{ Bern:2019nnu, Bern:2019crd, Bern:2020buy, Bern:2020uwk}. Advanced quantum amplitude techniques such as generalized unitarity \cite{ Bern_1994, Bern_1995} and double copy \cite{ Kawai:1985xq, PhysRevD.78.085011, PhysRevLett.105.061602} have been used to develop “dictionaries” that translate quantum scattering amplitudes to classical gravitational dynamics, supplementing PM calculations using effective field theory \cite{ Kalin:2020mvi, Kalin:2020fhe, Liu:2021zxr, Dlapa:2021npj, PhysRevLett.128.161104, Kalin_2023}.

A complementary method of calculating the scatter angle is provided by the self-force approximation, in which the scatter angle is expanded order-by-order in the mass ratio, assumed small, without recourse to the weak-field assumption present in PM theory. Thus self-force calculations are applicable even for strong-field orbits, assuming the mass ratio is small. Comparison with PM results in an overlapping domain of validity can provide useful mutual checks on both schemes. Furthermore, as pointed out by Damour in Ref.~\cite{ Damour2020}, the full conservative two-body dynamics, at any mass ratio, can be inferred through 4PM order using simply the leading-order self-force correction to the scatter angle. A calculation of the second-order self-force correction would extend this to 6PM order.

A first step towards a full self-force calculation of the scatter angle was taken in Ref.~\cite{LongBarack2021}, which demonstrated a method to reconstruct the linear metric perturbation sourced by a point mass moving along a fixed hyperbolic geodesic in the Schwarzschild spacetime, in a gauge suitable for self-force calculations. To demonstrate the practicality of their approach, they developed a time-domain (TD) numerical scheme for obtaining a certain scalar-like Hertz potential from which the metric perturbation can be derived. But that initial work stopped short of a calculation of the scatter angle correction itself. In a subsequent work 
by the same authors~\cite{BarackLong2022}, a first calculation of the self-force correction to the scatter angle was carried out, albeit in a scalar-field toy model. This numerical calculation was performed based on the time-domain computational platform developed in Ref.~\cite{LongBarack2021}. In addition to these numerical results, the scalar, electromagnetic and gravitational self-force corrections to the scatter angle have been derived analytically at leading (2PM) order, by Gralla and Lobo in Ref.~\cite{GrallaLobo2022}. These calculations were most recently extended to 4PM order using scattering-amplitudes methods, showing impressive agreement with the self-force results \cite{Barack:2023oqp}.

Time-domain codes provide the most straightforward route to begin self-force calculations along scatter orbits, but they are not necessarily optimal. Frequency-domain (FD) methods provide an alternative approach to self-force calculations. Their applications include first calculations of the scalar-field self-force along equatorial bound \cite{ Warburton:2010eq, Warburton:2011hp} and generic bound \cite{Nasipak:2019hxh} geodesics in the Kerr spacetime, and the first calculations of the gravitational self-force along generic bound geodesics in the Kerr spacetime \cite{vandeMeent2016, vandeMeent2018}. These methods are prized for their accuracy and efficiency, but so far they have only been applied to bound orbits. There is, therefore, natural interest in extending frequency-domain methods to the study of self-force along scatter orbits, a non-trivial task that presents several new obstacles, outlined below. A scalar-field toy model in the Schwarzschild spacetime provides the simplest laboratory in which to meaningfully explore and mitigate these challenges. The development and numerical implementation of such a model will be the focus of the current work.

A mainstay of frequency-domain self-force calculations is the so-called {\it method of extended homogeneous solutions} (EHS), which was introduced in Ref.~\cite{Barack:2008ms} to overcome the Gibbs phenomenon that would beset a naive attempt to reconstruct the time-domain field from frequency modes at the location of the particle. In this approach, the time-domain field is reconstructed from certain {\it homogeneous} frequency modes even inside the libration region of the particle source, ensuring a uniform exponential convergence of the Fourier mode sum, even at the particle. The applicability of this idea relies crucially on the libration region having a compact radial extent. This is the case for bound orbits, but, unfortunately, not for scattering orbits. 
(A frequency-domain method was used in \cite{Hopper:2017qus, Hopper2018}
to compute the gravitational radiation emitted by a point mass scattered off a Schwarzschild black hole, but, crucially, that work considered only asymptotic waveforms and fluxes, which did not require the use of EHS at all.)

To describe the situation more accurately, let the Schwarzschild radial coordinate of the particle be described by the function $r=r_p(\tau)$, where $\tau$ is the particle's proper time. The timelike worldline of the scattering orbits splits the exterior of the central Schwarzschild black hole into two disjoint vacuum regions, $2M<r<r_p(\tau)$ and $r>r_p(\tau)$--- the orbit's ``interior'' and ``exterior'', respectively. When the method of EHS is applied to bound orbits, a time-domain solution is constructed on either side of the orbit. In the case of a scattering orbit, a solution can only be constructed in this way in the orbit's interior, but not in its exterior; see Sec.~\ref{sec:EHS_section} for a discussion.
This poses a problem in practice, because the standard radiation-gauge approach to gravitational self-force calculations relies on a two-sided mode-sum regularization procedure, which requires field derivatives taken from both sides of the orbit. A radiation-gauge calculation based on one-side derivatives is much more complicated (and has not been attempted yet) \cite{Pound:2013faa}.  The issue does not occur in Lorenz-gauge calculations, where a one-sided mode-sum regularization is easy to apply \cite{Barack:2001gx}. However, frequency-domain Lorenz-gauge calculations are not yet as well developed, especially in Kerr. 

Fortunately, a one-sided mode-sum regularization is also easily applied in our scalar-field model, and this is the approach we adopt in this paper. That is, we use the EHS method to construct solutions (only) in the interior of the orbit, and from these we derive the self-force. 

A crucial step in applying EHS involves the calculation of certain normalization integrals $C_{\lmw}^-$ (one for each frequency-harmonic mode), which are obtained by numerically evaluating radial integrals along the orbit. These integrals now stretch to radial infinity, and exhibit slow, oscillatory convergence, rendering them slow to evaluate, and making the error from any finite-radius truncation hard to control. The problem is analysed in Sec.~\ref{sec:radial_integral}, and two complementary solutions are developed. The first, the \textit{tail correction scheme}, makes use of analytic expansions of the integrand at large radius to obtain an analytic approximation to the neglected tail of the integral. The second uses \textit{integration by parts} (IBP) to increase the decay rate of the integrand at large radius, reducing the truncation error for a given truncation radius. Evaluation of the normalization integrals is also accelerated by use of specialized quadrature rules and parallelization. 

The next step involves the reconstruction of the time-domain multipole modes of the scalar field from the frequency modes, at the location of the particle. These 
form the direct input to the mode-sum regularization procedure. 
In the bound-orbit case, the periodicity of the orbit results in a discrete spectrum for the scalar field, and values of $C_{\lmw}^-$ may readily be reused, avoiding duplicate calculations when evaluating different components of the self-force or evaluating at different orbital positions. In the case of unbound motion, however, the spectrum is continuous and we must evaluate a Fourier \textit{integral}. In general this requires more frequency modes to be calculated, exacerbating the aforementioned issues with the evaluation of $C_{\lmw}^-$. Also, how best to store and reuse $C_{\lmw}^-$ data becomes an additional issue to be addressed. We approach this by calculating discretized $C_{\lmw}^-$ data in advance, and then using interpolation to obtain the values at the intermediate frequencies required by the Fourier integration routine. 

The structure of this paper is as follows. In Sec.~\ref{sec:model} we introduce our model, describing hyperbolic geodesics in the Schwarzschild spacetime, reviewing the notion of scalar-field self-force, and deriving an expression for the frequency modes of the physical, inhomogeneous field sourced by a scalar charge moving along a hyperbolic geodesic. In Sec.~\ref{sec:SF_from_modes} we discuss the use of mode-sum regularization and the decomposition of the self-force into conservative and dissipative pieces. We then explain the challenge of applying the EHS method to unbound orbits. Section~\ref{sec:radial_integral} illustrates the truncation problem when evaluating the EHS normalization integrals, and introduces the correction and IBP approaches to resolve this. The numerical methods we use (including time-domain reconstruction) are summarised in Sec.~\ref{sec:numerical_method}. Section~\ref{sec:FD_results} displays sample $C_{\lmw}^-$ spectra, as calculated using our methods, and discusses their features and the effectiveness of the different approaches we developed. 

The results of our self-force calculations are then presented in Sec.~\ref{sec:SF_reconstruction}. Comparisons with analytically known regularization parameters are used to provide internal validation, bolstered by comparison with the results of the time-domain code developed in Refs.~\cite{LongBarack2021, BarackLong2022}. At this point we note the frequency-domain code's potential for achieving significantly greater accuracy at small radii, but also observe a distressing breakdown of the frequency-domain code at large-$\ell$ as we move outwards along the orbit. This breakdown is reduced to a more gradual loss of accuracy by use of dynamic $\ell$ truncation, but an example calculation of the scatter angle is used to illustrate the remaining limitations of our code at large radii. Section~\ref{sec:cancellation_problem} explains the origin of the large-$\ell$ breakdown in terms of a cancellation problem inherent to the EHS approach---the same problem first noted in Ref.~\cite{vandeMeent2016} in relation to high-eccentricity bound orbits---and some potential remedies are suggested. We conclude in Sec.~\ref{sec:discussion} with a discussion of the progress we have made, the remaining challenges, and the direction of our future work.

Throughout this paper we work in natural units with $G =1= c$ and metric signature $(-,+,+,+)$. The central object is represented by a background Schwarzschild spacetime of mass $M$, which, in Schwarzschild coordinates $x^{\alpha} = (t,r,\theta,\vf)$, has the line element
\begin{align}
    ds^2 = -f(r)dt^2 + f(r)^{-1}dr^2 + r^2d\Omega^2,
\end{align}
where $f(r) := 1 - 2M/r$, and $d\Omega^2 := d\theta^2 + \sin^2\theta\>d\vf^2$ is the metric on a unit 2-sphere. The Levi-Civita connection compatible with this metric is denoted $\nabla_{\mu}$.  The smaller object is described by a pointlike particle endowed with mass $\mu \ll M$ and ``small'' scalar charge $q \ll \sqrt{\mu M}$. Its trajectory is described by a worldline $x_p^{\alpha}(\tau)$ in the background spacetime, parameterized by proper time $\tau$, with 4-velocity $u^{\alpha}(\tau) := dx_p^{\alpha}/d\tau$.

\section{Scalar-charge Model}\label{sec:model} 

\subsection{Scattering geodesics in Schwarzschild spacetime}\label{sec:scatter_geodesics}

In the test particle limit $\mu/M \rightarrow 0$ and $q^2/(\mu M) \rightarrow 0$, the small object moves along a geodesic in the background Schwarzschild spacetime, which, without loss of generality, may be taken to lie in the equatorial plane, $\theta = \pi/2$ . The timelike and azimuthal Killing vectors give rise to conserved quantities $E$ (specific energy) and $L$ (specific angular momentum), respectively given by
\begin{align}
	 E &= f(r_p)\dot{t}_p\label{eq:tdot},\\
	 L &= r_p^2 \dot{\varphi}_p\label{eq:phidot},
\end{align}
where overdots denote derivatives with respect to proper time. The normalization of the 4-velocity, $u^{\alpha}u_{\alpha} = -1$, gives rise to an effective potential equation for the radial motion,
\begin{align}
	\dot{r}_p = \pm \sqrt{E^2 - V(r_p; L)} \label{eq:RadialEqn},
\end{align}
with effective potential
\begin{align}
	V(r; L) := f(r)\left(1+\frac{L^2}{r^2}\right).
\end{align}

We are interested in the scattering problem, in which $r_p\to\infty$ as $t\to\pm\infty$, which requires $E > 1$. Note that the 3-velocity at infinity, $v^i:= \frac{dx_p^i}{dt}\big|_{r\rightarrow\infty} \> (i = r, \theta, \phi)$, has magnitude
\begin{align}
	v &:= \sqrt{\left(v^r\right)^2 + \left( \left(rv^{\varphi}\right)_{\infty}\right)^2} = \frac{\sqrt{E^2-1}}{E},
\end{align}
leading to
\begin{align}	
	 E &= (1-v^2)^{-1/2},\label{eq:gamma_factor}
\end{align}
the standard Lorentz factor. The particle scatters back to infinity if, and only if, $L > L_{\rm crit}(E)$, where
\begin{align}
	L_{\rm crit}(E) = \frac{M}{Ev}\sqrt{(27E^4+9\nu E^3 - 36E^2 - 8\nu E+8)/2},
\end{align}
with $\nu := \sqrt{9E^2 - 8}$.

We may use the first integrals $E$ and $L$ to parameterize our orbit, or we may choose to replace $L$ with the \textit{impact parameter} $b$:
\begin{align}
	b := \lim_{\tau\to-\infty} r_p(\tau)\sin|\varphi_p(\tau)-\varphi_p(-\infty)| = \frac{L}{\sqrt{E^2-1}}.\label{eq:bdef}
\end{align}	
Likewise, we may replace $E$ with $v$ using Eq.~\eqref{eq:gamma_factor}. The orbit $(b, v)$ is then a scatter orbit provided
\begin{align}
	b > b_{\rm crit}(E) := \frac{L_{\rm crit}(E)}{\sqrt{E^2 - 1}}.
\end{align}

For a given $E > 1$ and $L > L_{\rm crit}(E)$, the cubic equation 
\begin{align}
    \dot{r}_p^2 = E^2 - V(r;L) = 0\label{eq:turning_points}
\end{align} has three real roots $r_1, r_2$ and $\rmin$, with $ r_1 < 0$ and $2M < r_2 < r_{\rm min}$.  These are given explicitly by \cite{maartenUnpublished}
\begin{align}
    r_1 &= \frac{6M}{1 - 2\zeta \sin\left(\frac{\pi}{6}+\xi\right)}, \label{eq:r1_explicit}\\
    r_2 &= \frac{6M}{1+2\zeta\cos\xi}, \label{eq:r2_explicit}\\
    \rmin &= \frac{6M}{1-2\zeta\sin\left(\frac{\pi}{6}-\xi\right)}, \label{eq:rmin_explicit}
\end{align}
where $\zeta := \sqrt{1-12M^2/L^2}$ and
\begin{align}
    \xi := \frac{1}{3}\arccos{\left(\frac{1+(36-54E^2)M^2/L^2}{\zeta^3}\right)}.
\end{align}
The root $\rmin$ is the periastron radius for the orbit, and may be calculated from $E$ and $L$ using Eq.~\eqref{eq:rmin_explicit}. At the same time, Eq.~\eqref{eq:turning_points} allows one to determine $L$ for a given $\rmin$ and $E > 1$. The pair $(E, r_{\rm min})$ provides an alternative parameterization for the orbit.

The orbital turning points also give rise to another parameterization, in terms of an eccentricity $e > 1$ and a semi-latus rectum $p$, defined by the relations
\begin{align}
    r_1 = \frac{Mp}{1-e}, \quad r_{\rm min} = \frac{Mp}{1+e}.\label{eq:ep_def}
\end{align}
Substituting Eqs.~\eqref{eq:ep_def} into Eq.~\eqref{eq:turning_points} and solving for $E$ and $L$, one finds the same relations as in the bound case,
\begin{align}
	E^2 = \frac{(p-2)^2 - 4e^2}{p(p-3-e^2)}, \quad\quad L^2 = \frac{p^2M^2}{p-3-e^2}. \label{eq:ELepRelation}
\end{align}
To invert Eqs.~\eqref{eq:ELepRelation}, it is easiest to first use Eq.~\eqref{eq:ep_def} to write $p$ in terms of $\rmin$ and $e$, and then solve the second equation in \eqref{eq:ELepRelation} to find
\begin{align}
    e = &\frac{L^2\rmin - 2M\rmin^2}{2M(L^2 + \rmin^2)}\\ \nonumber  &\quad+\frac{\sqrt{L^4(\rmin^2 + 4M\rmin - 12M^2) - 16L^2M^2\rmin^2} }{2M(L^2 + \rmin^2)}.
\end{align}
Using Eq.~\eqref{eq:rmin_explicit}, we thus get $e(E, L)$, and hence also $p(E,L) = \rmin(1+e)/M$ using Eq.~\eqref{eq:ep_def}. 

With the $(e, p)$ parameterization, the radial motion is described in the familiar Keplerian-like form,
\begin{align}
    r_p(\chi) = \frac{Mp}{1+e\cos\chi}, \label{eq:epparam}
\end{align}
in terms of the relativistic anomaly $\chi$. This anomaly takes values in $-\chi_{\infty} < \chi < \chi_{\infty}$, where $\chi_{\infty} := \arccos{(-1/e)}$ corresponds to the particle returning to infinity, and $\chi = 0$ corresponds to the periastron passage.
The $(e,p)$ parameterization is also convenient for calculating the other components of $x_p^{\alpha}$, using $\chi$ as the parameter along the orbit. $t_p(\chi)$ can be obtained using Eqs.~\eqref{eq:tdot} and \eqref{eq:RadialEqn}, and then substituting Eqs.~\eqref{eq:ELepRelation} and \eqref{eq:epparam}:
\begin{align}
    \frac{dt_p}{d\chi} =\nonumber \frac{\dot{t}_p}{\dot{r}_p}\frac{dr_p}{d\chi}= &\frac{Mp^2}{(p-2-2e\cos\chi)(1+e\cos\chi)^2}\\
    &\quad\times\sqrt{\frac{(p-2)^2-4e^2}{p-6-2e\cos\chi}}.\label{eq:dt_dchi}
\end{align}
This equation can then be integrated numerically, subject to an initial condition, to give $t_p(\chi)$. In this work we chose to take $t_p = 0$ at periastron ($\chi = 0$), which gives the symmetry relation $t_p(-\chi) = -t_p(\chi)$. Similarly, we express
\begin{align}
	\frac{d\vf_p}{d\chi} = \frac{\dot{\vf}_p}{\dot{r}_p}\frac{dr_p}{d\chi},
\end{align}
which, using Eqs.~\eqref{eq:phidot} and \eqref{eq:RadialEqn} and then substituting from Eqs.~\eqref{eq:ELepRelation} and \eqref{eq:epparam}, gives
\begin{align}
	\frac{d\vf_p}{d\chi} =  \sqrt{\frac{p}{p-6-2e\cos \chi}}\label{eq:dphidchi}.
\end{align}
We can integrate this up to give
\begin{align}
    \vf_p(\chi) &= 2\sqrt{\frac{p}{p-6-2e}}\displaystyle\int_0^{\chi/2} \frac{d\theta}{\sqrt{1+k^2\sin^2\theta}} \\
						&= k\sqrt{\frac{p}{e}}\El_1\left({\frac{\chi}{2}, -k^2}\right), \label{eq:ellphichi}
\end{align}
where $k^2 := 4e/(p-6-2e)$, and $\El_1(\phi, z)$ is the incomplete elliptic integral of the first kind with parameter $z$: 
\begin{align}
	\El_1(\phi, z) = \displaystyle\int_0^{\phi} \frac{d\theta}{\sqrt{1-z\sin^2 \theta}}.
\end{align}
Note that we selected the initial condition $\vf_p(\chi=0) = 0$, which once again gives rise to a symmetry, $\vf_p(-\chi) = -\vf_p(\chi)$.

We define $\varphi_{\rm in}$ and $\varphi_{\rm out}$ to be the asymptotic values of $\varphi_p$ as $\chi\rightarrow -\chi_{\infty}$ and $\chi\rightarrow\chi_{\infty}$ respectively. The \textit{scatter angle} is then defined to be
\begin{align}
	\delta\vf := \vf_{\rm out} - \vf_{\rm in} - \pi,
\end{align}
which for a geodesic trajectory is given by
\begin{align}
	\delta\vf = 2k\sqrt{\frac{p}{e}}\El_1\left(\frac{\chi_{\infty}}{2}, -k^2\right) - \pi.\label{eq:geodesic_scatter_angle}
\end{align}
Here we used Eq.~\eqref{eq:ellphichi} along with the identity $\El_1(-\phi, m) = - \El_1(\phi, m)$.

Finally, using $r$ as a parameter along the outbound leg of the orbit ($\dot{r}_p > 0$), the relations $t_p(r)$ and $\varphi_p(r)$ admit useful large-$r$ expansions in $1/r$. For example, for $t_p$ we find
\begin{align}
	t_p(r) = t_0 + \frac{r}{v} + 2MB\log\left(\frac{r}{2M}\right) + 2M\displaystyle\sum_{n=1}^{\infty}C_n\left(\frac{2M}{r}\right)^n\label{eq:texp}
\end{align}
as $r \rightarrow\infty$, where the constants $B$ and $C_n$ are given analytically in terms of $E$ and $L$ in Appendix~\ref{app:geodesic_expansion_coeffs} for $n \leq 5$. The constant $t_0$ is fixed by the boundary condition $t_p(\rmin) = 0$. Likewise, 
\begin{align}
	\varphi_p(r) = \varphi_{\rm out} + \displaystyle\sum_{n=1}^{\infty}D_n\left(\frac{2M}{r}\right)^n\label{eq:phiexp}
\end{align}
as $r\rightarrow\infty$. 
The constants $D_n$ are given analytically in Appendix~\ref{app:geodesic_expansion_coeffs}. Expressions along the inbound leg of the orbit may be obtained by using the symmetry relations $t_p(\chi) = -t_p(-\chi)$ and $\varphi_p(\chi) = -\varphi_p(-\chi)$. We will also make use of the large-$r$ expansion for the radial component of the 4-velocity,
\begin{align}
	\SP{\frac{dr_p}{d\tau}}^{-1} =\sum_{n=0}^{\infty}U_n\left(\frac{2M}{r}\right)^n,\label{eq:urinv_exp}
\end{align}
as $r \rightarrow\infty$, where the first few coefficients $U_n$ are given in Appendix~\ref{app:geodesic_expansion_coeffs}. 

\begin{figure}
  \centering
  \includegraphics[width=0.25\textwidth]{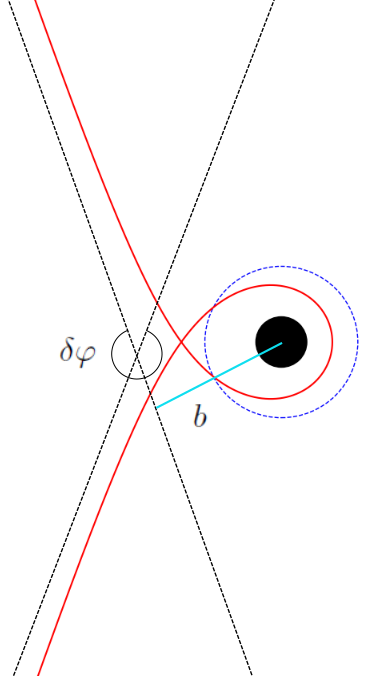}~\\~\\
  \caption[Geometric meaning of the impact parameter and scatter angle for hyperbolic orbits]{\label{fig:impact_and_scatter}Geometric interpretation of the impact parameter $b$ and the scatter angle $\delta\varphi$ (modulo $2\pi$). The geodesic orbit displayed here has $E = 1.1$ and $r_{\rm min} = 4M$, corresponding to $L \approx 4.7666M$, $b \approx 10.4015M$,  $v\approx 0.4166$, $e \approx 1.6273$ and $p \approx 10.5092$. The scatter angle is $\delta\vf \approx 323^{\circ}$. The view is in the equatorial plane, plotted on axes $x = r\cos\varphi$ and $y =r\sin\varphi$. The black hole (black disk) and the innermost stable circular orbit (blue circle) are to scale.} 
\end{figure}

Above we have introduced several alternative parametrizations for scattering geodesics. In this paper we will primarily use $(E, \rmin)$. This choice is convenient, because it allows one to control how relativistic the particle motion is at infinity (by varying $E$), and also how deep the orbit penetrates into the strong-field region (by varying $\rmin$). Once we have specified $(E, \rmin)$, we will then also calculate and use the corresponding values of $L$, $b$, $v$, $e$ and $p$. A sample scattering orbit, with parameters $E = 1.1$ and $\rmin = 4M$, is depicted in Fig.~\ref{fig:impact_and_scatter}. 

\subsection{Scalar-field self force}\label{subsec:SSF_intro}

The particle sources a scalar field $\Phi$, which we assume is massless and minimally coupled. This field obeys the Klein-Gordon equation on the background Schwarzschild spacetime, 
\begin{align}
	\nabla_{\mu}\nabla^{\mu}\Phi = -4\pi T, \label{eq:ScalarEOM}
\end{align}
where the scalar charge density is
\begin{align}
	T\left(x^{\alpha}\right) := q\displaystyle\int_{-\infty}^{+\infty} \delta^{4}\left(x^{\alpha} - x_p^{\alpha}(\tau)\right)\frac{d\tau}{\sqrt{-g(x)}}, \label{eq:ScalarSource}
\end{align}
with $g$ being the determinant of the Schwarzschild metric. 

The full (retarded) scalar field may be decomposed as
\begin{align}
	\Phi = \Phi^R + \Phi^S,\label{eq:DetWhitDecomp}
\end{align} 
where $\Phi^R$ and $\Phi^S$ are the Detweiler-Whiting regular and singular fields respectively, introduced in \cite{DetweilerWhiting2003}. The regular field $\Phi^R$ is a certain vacuum solution to the scalar field equation, smooth everywhere, including at the particle's location. The singular field $\Phi^S$ is a particular solution of (\ref{eq:ScalarEOM}), singular at the worldline. Interaction with its own scalar field modifies the particle's trajectory according to \cite{Quinn00}
\begin{align}
	u^{\nu}\nabla_{\nu}\left(\mu u^{\mu}\right) = q\nabla^{\mu}\Phi^R.\label{eq:scalar_self_forced_EOM}
\end{align}
 Note that we have not included the gravitational self-force acting on the particle, nor the effect of the scalar field's backreaction on the background spacetime. 

We refer to the quantity on the right-hand side of Eq.~\eqref{eq:scalar_self_forced_EOM} as the \textit{scalar-field self-force}, 
\begin{align}
	F_{\text{self}}^{\alpha} := q\nabla^{\alpha}\Phi^R \propto q^2.\label{eq:SSF_def}
\end{align} 
The singular field $\Phi^S$ does not appear; the self-force arises due to the interaction between the particle and its regular field only. Furthermore, the derivative on the right-hand side is not generically orthogonal to the 4-velocity, $u^{\mu}$. The implications of this become clear if one splits Eq.~\eqref{eq:scalar_self_forced_EOM} into parts parallel and perpendicular to $u^{\alpha}$:
\begin{align}
	\frac{d\mu}{d\tau} &= - qu_{\mu}\nabla^{\mu}\Phi^R,\label{eq:SSF_EOM_par}\\
	\mu u^{\nu}\nabla_{\nu}u^{\mu} &= q\left(g^{\mu\nu} + u^{\mu}u^{\nu}\right)\nabla_{\nu}\Phi^R\label{eq:SSF_EOM_perp}.
\end{align}
From this we see that the component parallel to $u^{\alpha}$ is responsible for a variation in the particle's rest mass. Equation~\eqref{eq:SSF_EOM_par} can be integrated to give
\begin{align}
    \mu(\tau) = \mu_0 - q\Phi^R(\tau),
\end{align}
where $\mu_0$ is a constant of integration. It is expected that $\Phi^R(-\infty) = \Phi^R(\infty)$, in which case there is no net mass change overall. Equation~\eqref{eq:SSF_EOM_perp}, meanwhile, can be rewritten as
\begin{align}
    u^{\nu}\nabla_{\nu}u^{\mu} &= \eta_q F^{\mu}_{\bot}\label{eq:SSF_EOM_perp_alt},
\end{align}
where $\eta_q := q^2/(\mu M) \ll 1$, and the perpendicular components of the self-force are defined by
\begin{align}
    \eta_q F_{\bot}^{\mu} &:= \frac{q}{\mu}\left(g^{\mu\nu} + u^{\mu}u^{\nu}\right)\nabla_{\nu}\Phi^R.\label{eq:SSF_perp_def}
\end{align}
$F_{\bot}^{\mu}$ gives rise to the self-acceleration that alters the trajectory. 

In particular, it was shown in Ref.~\cite{BarackLong2022} that the scattering angle can be expanded in the form
\begin{align}
    \delta\vf = \delta\vf^{(0)} + \eta_q\delta\vf^{(1)} + O(\eta_q^2),\label{eq:scatter_angle_expansion}
\end{align}
where $\delta\vf^{(0)}$ is the geodesic scatter angle given in Eq.~\eqref{eq:geodesic_scatter_angle}, and 
\begin{align}
    \delta\vf^{(1)} = \int_{-\chiinf}^{\chiinf} \SB{\mathcal{G}_E(\chi)F_t^{\bot}(\chi) - \mathcal{G}_L(\chi)F_{\varphi}^{\bot}(\chi)}\tau_{\chi}d\chi\label{eq:1SF_scatter_angle}
\end{align}
is the first-order self-force correction to the scatter angle. In Eq.~\eqref{eq:1SF_scatter_angle}, $\tau_{\chi} := dt/d\chi$ is evaluated along the background geodesic, $F_{\alpha}^{\perp}$ is the self-force that would be felt by a particle moving along the background geodesic, and the form of the functions $\mathcal{G}_{E/L}(\chi)$ may be found in Ref.~\cite{BarackLong2022}. 

The relations between orbital parameters derived in Sec.~\ref{sec:scatter_geodesics} are only valid for geodesic orbits, so when expanding the self-force as in Eq.~\eqref{eq:scatter_angle_expansion}, it is important to be clear which pair of orbital parameters are being taken to be fixed. Fixing different choices of parameters yields different values of $\delta\vf^{(1)}$. In Ref.~\cite{BarackLong2022}, the parameters $(b, v)$ are taken to be fixed, and Eq.~\eqref{eq:1SF_scatter_angle} is correct for this convention. The pair $(b, v)$ are useful parameters in this context, because they are defined in terms of properties of the orbit at infinity, which (in the analogous gravitational problem) removes gauge ambiguities. We shall continue to use $(E, \rmin)$ as convenient parameters to describe geodesic orbits. However, when we perform our calculation of the scatter angle in Sec.~\ref{sec:scatter_angle_calc}, we will \textit{not} be calculating $\delta\vf^{(1)}$ with fixed $(E, \rmin)$. Instead, we will be calculating with fixed values of $(b, v)$ equal to those of the \textit{geodesic} with parameters $(E, \rmin)$.



\subsection{Mode decomposition}
As a first step towards the solution of the scalar field equation \eqref{eq:ScalarEOM}, the scalar field and scalar charge density are decomposed into a basis of spherical harmonics $Y_{\ell m}(\theta, \vf)$ defined on surfaces of constant $t$ and $r$ around the central black hole,
\begin{align}
	\Phi &= \sum_{\ell m} \frac{1}{r}\psi_{lm}(t, r) Y_{\ell m}(\theta, \varphi)\label{eq:spherharmPhi},\\
	T &= \sum_{\ell m} T_{\ell m}(t,r)Y_{\ell m}(\theta, \varphi).\label{eq:spherharmT}
\end{align}
Equation~\eqref{eq:ScalarEOM} then becomes
\begin{align}
	-\frac{\partial^2\psi_{\lm}}{\partial t^2}+\frac{\partial^2\psi_{\lm}}{\partial r_{*}^2} - V_l(r)\psi_{\lm}= -4\pi rf(r) T_{\lm}, \label{eq:ScalarEOMTD}
\end{align}
where $r_* := r + 2M\log\SP{\frac{r}{2M}-1}$ is the Regge-Wheeler tortoise coordinate, and the potential $V_{\ell}(r)$ is defined by
\begin{align}
    V_{\ell}(r) := \left(\frac{l(l+1)}{r^2}+\frac{2M}{r^3}\right)f(r) \label{eq:ScalarEffPot}.
\end{align}
As a final step, we make a Fourier decomposition
\begin{align}
	\psi_{\lm}(t,r) &= \displaystyle\int_{-\infty}^{+\infty}d\omega\, e^{-i\omega t}\psi_{\lm\omega}(r) \label{eq:freq_decomp_field},\\
	T_{\lm}(t,r) &= \displaystyle\int_{-\infty}^{+\infty}d\omega \, e^{-i\omega t}T_{\lm\omega}(r),
\end{align}
to get the frequency-domain radial equation
\begin{align}
	\frac{d^2\psi_{\lm\omega}}{dr_{*}^2} - (V_l(r)-\omega^2)\psi_{\lm\omega} = -4\pi rf(r)T_{\lm\omega}. \label{eq:ScalarEOMFD}
\end{align}

The source modes $T_{\lm\omega}$ are obtained as follows. First, integrating in Eq.~\eqref{eq:ScalarSource}, we obtain
\begin{align}
	T\left(x^{\alpha}\right) = \frac{q}{r_p^2 u^t}\delta(r-r_p(t))\delta(\varphi - \varphi_p(t))\delta(\theta-\pi/2). \label{eq:ScalarSourceResolved}
\end{align}
The spherical harmonics take the form $Y_{\lm}(\theta,\varphi) = c_{\lm}e^{im\varphi}P_{\lm}(\cos\theta)$, where $c_{\lm}$ are certain real constants and $P_{\lm}$ are the associated Legendre polynomials. Using the orthogonality relations we have
\begin{align}
	T_{\lm}(t,r) &= \displaystyle\int d^2\Omega \>Y^{*}_{\lm}(\theta,\varphi) T(t,r,\theta,\vf),
\end{align}
where $*$ denotes complex conjugation, and substituting for $T$ from Eq.~\eqref{eq:ScalarSourceResolved}, we obtain
\begin{align}
    T_{\lm}(t,r) &= d_{\lm}e^{-im\varphi_p(t)}\frac{q}{r_p^2 u^t}\delta(r-r_p(t)),
\end{align}
where $d_{\lm} := c_{\lm}P_{\lm}(0)$ is a constant. A Fourier transform now yields
\begin{align}
	T_{\lm\omega} = \frac{1}{2\pi}\displaystyle\int_{-\infty}^{+\infty}dt\> d_{\lm} e^{-im\varphi_p(t)}\frac{q}{r_p^2u^t}\delta(r-r_p(t))e^{i\omega t}.
\end{align}
Switching integration variable to $r_p$ and using the orbital symmetries $t_p(-\chi) = -t_p(\chi)$ and $\vf_p(-\chi) = -\vf_p(\chi)$, gives
\begin{align}
	T_{\lm\omega}(r) &= \frac{qd_{\lm}}{\pi}\displaystyle\int_{r_{\rm min}}^{+\infty}\frac{dr_p}{|\dot{r}_p|} \frac{\delta(r-r_p)}{r_p^2}\cos\SP{\omega t_p(r_p)-m\varphi_p(r_p)}\\
	&= \frac{qd_{\lm}}{\pi r^2 |\dot{r}_p(r)|}\cos\left(\omega t_p(r) - m\varphi_p(r)\right)\Theta(r-r_{\rm min}) \label{eq:SourceFT}.
\end{align}

Finally, we note that conjugation symmetry relates some modes to others. Since $\Phi$ is a real scalar field, we have that $\Phi^* = \Phi$ and hence, using Eq.~\eqref{eq:spherharmPhi},
\begin{align}
	\Phi = \sum_{\lm}\frac{1}{r}\psi_{\lm}^*Y_{\lm}^*.
\end{align}
Recalling the identity
\begin{align}
	Y_{\lm}^*(\theta, \varphi) = (-1)^mY_{\ell,-m}(\theta, \varphi),
\end{align}
we may rewrite $\Phi$ as
\begin{align}
	\Phi = \sum_{\lm}\frac{1}{r}(-1)^m\psi_{\lm}^*Y_{\ell,-m},
\end{align}
and hence obtain
\begin{align}
	\psi_{\lm}(t,r) = (-1)^m\psi_{\ell,-m}^*(t,r).\label{eq:negmreln}
\end{align}
This means we only need to calculate modes with $m \geq 0$. Furthermore, when $\ell + m$ is odd, $d_{\lm} = 0$ and hence the source $T_{\lm}$ vanishes. From this we conclude that the modes of the retarded field with odd $\ell + m$ are identically zero, everywhere.  The $\theta$-derivative of the scalar field also vanishes on the equator by symmetry. Thus, for a given $\ell$-mode we only need to calculate those modes with $m \geq 0$ and $\ell+m$ even, roughly one quarter as many as naively expected.  


\subsection{Homogeneous solutions}\label{sec:homsols}
We first consider the solutions to the homogeneous form of Eq.~\eqref{eq:ScalarEOMFD},
\begin{align}
	\frac{d^2\psi_{\lm\omega}}{dr_{*}^2} - (V_l(r)-\omega^2)\psi_{\lm\omega} = 0. \label{eq:ScalarEOMFD_hom}
\end{align}
From  Eq.~\eqref{eq:ScalarEffPot}, we see that $V_{\ell}(r) \rightarrow 0$ as $r_* \rightarrow \pm\infty$, i.e. at the horizon and infinity. In those limits, the radial equation reduces to a harmonic oscillator,
\begin{align}
    \frac{d^2\psi_{\lm\omega}}{dr_{*}^2} +\omega^2\psi_{\lm\omega} \approx 0,
\end{align}
whose solution is given by a superposition of sinusoidal modes $e^{i\omega r_*}$ and $e^{-i\omega r_*}$. 

The physical, inhomogeneous, scalar field sourced by the particle should obey retarded boundary conditions with purely ingoing radiation at the horizon, and asymptotically outgoing radiation at infinity. In the frequency domain, using the Fourier conventions of Eq.~\eqref{eq:freq_decomp_field}, this requirement translates to
\begin{align}
	\psi_{lm\omega} &\sim e^{i\omega r_{*}} \>\>\>\> \text{  as } r_{*}\rightarrow +\infty  \label{eq:BCinfty},\\
	\psi_{lm\omega} &\sim e^{-i\omega r_{*}}\>\>\text{  as } r_{*} \rightarrow -\infty \label{eq:BChorizon}.
\end{align}
It is therefore convenient to define the basis of homogeneous solutions $\{\psi_{\ell\omega}^-, \psi_{\ell\omega}^+\}$, which for $\omega \neq 0$ are defined to be the solutions to the homogeneous equation~\eqref{eq:ScalarEOMFD_hom} obeying the boundary conditions 
\begin{align}
    \psi_{\ell\omega}^{\pm} \sim e^{\pm i\omega r_*} \>\>\>\> \text{   as } r_* \rightarrow\pm\infty.\label{eq:def_homsols}
\end{align}
We note that neither these  boundary conditions nor the homogeneous equation \eqref{eq:ScalarEOMFD_hom} depend on the mode number $m$, so that the homogeneous solutions $\psi_{\ell\omega}^{\pm}(r)$ can be labelled only by $\ell$ and $\omega$. 

These homogeneous solutions may be expanded as a series in the appropriate wave zone. For example, as $r \rightarrow \infty$, we have
\begin{align}
	\psi_{\lw}^+(r) &= e^{i\omega r_*}\sum_{k = 0}^{k_{\rm out}} c_k^{\infty}\left(\frac{2M}{r}\right)^{k}  + O\SP{\frac{2M}{r} }^{k_{\rm out}+1},\label{eq:asymp_inf}
\end{align}
where the coefficients $c_{k >0 }^{\infty}$ depend on $\ell$ and $\omega$, and  are determined in terms of $c_0^{\infty}$ using a recurrence relation described in Appendix~\ref{app:homsol_BCs}. We adopt an overall normalization such that $c_0^{\infty}=1$. Likewise, in the near-horizon wave zone, $r \rightarrow 2M$, we have an expansion 
\begin{align}
 \psi_{\lw}^-(r) &= e^{-i\omega r_*}\sum_{k = 0}^{k_{\rm in}} c_k^{\text{eh}}\SP{\frac{r}{2M}-1}^k + O\SP{r-2M}^{k_{\rm in}+1}, \label{eq:asymp_eh}
\end{align}
The coefficients $c_{k>0}^{\rm eh}$ are determined from $c_{0}^{\rm eh}$ using another recurrence relation, also summarised in Appendix~\ref{app:homsol_BCs}. We choose a normalization such that $c_0^{eh} = 1$. 

For $\omega = 0$, Eq.~\eqref{eq:ScalarEOMFD_hom} can be rewritten in the form
\begin{align}
	\frac{d}{d\rho}\left[(1-\rho^2)\frac{dR_{\ell}}{d\rho}\right] + \ell(\ell+1)R_{\ell} = 0,\label{eq:LegendreEqn}
\end{align}
where $R_{\ell} := \psi_{\ell 0}/r$ and $\rho = (r-M)/M$. The general solution is
\begin{align}
    R_{\ell}(\rho) = a_{\ell} P_{\ell}(\rho) + b_{\ell} Q_{\ell}(\rho),
\end{align}
for arbitrary constants $a_{\ell}$ and $b_{\ell}$. Here $P_{\ell}(\rho)$ is the Legendre polynomial, which is regular at all finite points but blows up as $\rho \rightarrow \pm\infty$ for $\ell > 0$, and $Q_{\ell}(\rho)$ is the Legendre function of the second kind, which decays at infinity but is singular at $\rho=1$ ($r=2M$). Thus for $\omega = 0$ we take our basis of homogeneous solutions to be 
\begin{align}
    \psi_{\ell0}^-(r) &:= r P_{\ell}\SP{\frac{r-M}{M}},\label{eq:psi_minus_def} \\
    \psi_{\ell0}^+(r) &:= r Q_{\ell}\SP{\frac{r-M}{M}}.
\end{align}
The large-$r$ behavior of these solutions, needed for later discussion, is 
\begin{align}
    \psi_{\ell0}^-(r) \sim r^{\ell+1},\quad\quad\psi_{\ell0}^+(r) \sim r^{-\ell}.\label{eq:static_homsol_asymp}
\end{align}


\subsection{The inhomogeneous solution}\label{sec:inhom_solution}
Solutions to the inhomogeneous frequency-domain equation \eqref{eq:ScalarEOMFD} can be found using variation of parameters. One such solution is given by
\begin{align}
    	\psi_{\lm\omega}(r) =  \>&\psi_{\lw}^+(r)\displaystyle\int_{r_{\rm min}}^r\frac{\psi_{\lw}^-(r')S_{\lm\omega}(r')}{W_{\lw}}\frac{dr'}{f(r')}\label{eq:VoPRadialEqnScattering} \\\nonumber &+ \psi_{\lw}^-(r)\displaystyle\int_{r}^{+\infty}\frac{\psi_{\lw}^+(r')S_{\lm\omega}(r')}{W_{\lw}}\frac{dr'}{f(r')}, 
\end{align}
where $	S_{\lm\omega} := -4\pi rf(r)T_{\lm\omega}$
is the source on the right-hand side of Eq.~\eqref{eq:ScalarEOMFD}, and $W_{\lw}:=\psi_{\lw}^-\frac{d\psi_{\lw}^+}{dr_*}- \psi_{\lw}^+\frac{d\psi_{\lw}^-}{dr_*}$ is the Wronskian of the homogeneous solutions, which depends only on $\ell$ and $\omega$, and not on $r$. For convenience we give names to the integrals in Eq.~\eqref{eq:VoPRadialEqnScattering}:
\begin{align}
	c_{\lm\omega}^+(r) &:= \displaystyle\int_{r_{\rm min}}^r\frac{\psi_{\lw}^-(r')S_{\lm\omega}(r')}{W_{\lw}}\frac{dr'}{f(r')} \label{eq:cplus},\\
	c_{\lm\omega}^-(r) &:= \displaystyle\int_{r}^{+\infty}\frac{\psi_{\lw}^+(r')S_{\lm\omega}(r')}{W_{\lw}}\frac{dr'}{f(r')}\label{eq:cminus}.
\end{align}

We will find that Eq.~\eqref{eq:VoPRadialEqnScattering} gives the correct retarded solution to Eq.~\eqref{eq:ScalarEOMFD}, except for the special case $\ell = 0 = \omega$ discussed below.  We show this for the non-static $\omega \neq 0$ modes first, with the first task being to show that the integral defining $c_{\lm\omega}^-$ converges. Substituting $T_{\lm\omega}$ from Eq.~\eqref{eq:SourceFT}, and recalling the expansions~\eqref{eq:texp}-\eqref{eq:urinv_exp} and \eqref{eq:def_homsols}, the integrand of $c_{\lm\omega}^-$ takes the schematic form
\begin{align}
	J_{\lm\omega}^-(r) \sim e^{i\omega(1+1/v)r}r^{i\omega(1+B)-1} + e^{i\omega(1-1/v)r}r^{i\omega(1-B)-1}
\end{align}
as $r \rightarrow\infty$, where $B$ is one of the constants appearing in expansion \eqref{eq:texp}.  The integral defining $c_{\lm\omega}^-$ thus converges like $\text{sinusoidal oscillations}/r$ at large radius. As we will see later, this slow oscillatory convergence is numerically  challenging.

Next we check the boundary conditions. Since the source is supported only on $r \geq r_{\rm min}$, the integral $c_{lm\omega}^+(r)$ vanishes for $r \leq r_{\rm min}$.  Thus, for $2M < r < r_{\rm min}$ we have
\begin{align}
	\psi_{\lm\omega}(r) &= C_{\lm\omega}^-\psi_{\lw}^-(r),
\end{align}
where we defined the normalization integral
\begin{align}
	C_{\lm\omega}^- := \displaystyle\int_{r_{\rm min}}^{+\infty}\frac{\psi_{\lw}^+(r')S_{\lm\omega}(r')}{W_{\lw}f(r')}dr'.\label{eq:Cminus_def}
\end{align}
Hence, as $r \longrightarrow 2M$,
\begin{align}
	 \psi_{\lm\omega}(r)	&\sim C_{\lm\omega}^-e^{-i\omega r_*},
\end{align}
as required. 

Meanwhile, as $r \longrightarrow \infty$,
\begin{align}
	\psi_{lw}^-(r) \sim a_{\lw}\>e^{i\omega r_*} + b_{\lw}\>e^{-i\omega r_*},
\end{align}
for some constants $a_{\lw}$ and $b_{\lw}$, and in particular it is bounded.
Since $c_{\lm\omega}^-(r) = O\SP{r^{-1}}$ as $r \rightarrow\infty$,
we have that
\begin{align}
	\lim_{r\rightarrow\infty} c_{\lmw}^-(r)\psi_{\lmw}^-(r) = 0.
\end{align}
Furthermore, as $r\to \infty$, $\psi_{\lw}^+(r) \sim e^{i\omega r_*}$ and thus
\begin{align}
	\psi_{\lmw}(r) \sim C_{\lmw}^+ e^{i\omega r_*},
\end{align}
where
\begin{align}
	C_{\lmw}^+ := \displaystyle\int_{r_{\rm min}}^{+\infty}\frac{\psi_{\lw}^-(r')S_{\lm\omega}(r')}{W_{\lw}f(r')}dr'.
\end{align}
This integral converges by a similar argument to that used for the convergence of $c_{\lmw}^-(r)$. Thus we have shown that the solution \eqref{eq:VoPRadialEqnScattering} is the retarded inhomogeneous solution for the non-static modes.

The situation is more subtle for the static modes $\omega = 0$. As before we must first check that the integral defining $c_{\lm\omega}^-$ is convergent and well-defined. The integrand of $c_{\lmw}^-(r)$ now takes the form
\begin{align}
    J_{\lm}(r) = \psi_{\ell0}^+(r)\cos(m\vf_p(r))\{r|\dot{r}_p(r)|\}^{-1},
\end{align}
where we have neglected overall numerical factors. Using Eq.~\eqref{eq:static_homsol_asymp}, we  find that the integrand goes like $r^{-(\ell+1)}$ as $r\rightarrow\infty$, and does not oscillate. That is because $\varphi_p(r)$ tends to a finite limit, and $\psi_{\ell0}^+$ is no longer asymptotically oscillatory. This means our integral converges for $\ell >0$, but not for $\ell = 0$. The form of Eq.~\eqref{eq:VoPRadialEqnScattering} will require modification for $\ell = 0$, and we will give further consideration to the large-$r$ behaviour of the static $\ell = 0$ solution below. For now, we return to the task of checking the boundary conditions for $\ell > 0$ and $\omega = 0$. 

When considering bound particle motion, the appropriate boundary condition for the static field is for it to be regular on the horizon, $r = 2M$, and decaying as $r \rightarrow \infty$. However, we find that these conditions cannot be imposed when the particle moves along a hyperbolic orbit, and instead we impose a condition of ``greatest regularity'', described below. 

The horizon boundary condition is easiest to investigate: because the source is supported only on $r \geq \rmin$, the field is given by
\begin{align}
    \psi_{\lm0}(r) = r C_{\lm0}^-P_{\ell}\SP{\frac{r-M}{M}}
\end{align}
for $2M  < r \leq \rmin$, where $C_{\lm0}^-$ is as defined in Eq,~\eqref{eq:Cminus_def} and we have substituted for $\psi_{\ell0}^-$ from Eq.~\eqref{eq:psi_minus_def}. The solution in Eq.~\eqref{eq:VoPRadialEqnScattering} is therefore regular at the horizon.

The large-$r$ behaviour is more subtle, because the factors $c_{\lm0}^+(r)$ and $\psi_{\ell0}^-(r)$ grow as $r\rightarrow\infty$, while $c_{\lm0}^-(r)$ and $\psi_{\ell0}^+(r)$ decay. This means that each term in Eq.~\eqref{eq:VoPRadialEqnScattering} consists of a decaying factor and a growing factor. A careful analysis using Eq.~\eqref{eq:static_homsol_asymp} gives that, for $\ell > 0$,
\begin{align}
    c_{\lm0}^{\pm}(r)\psi_{\lm0}^{\pm} (r)\sim r \quad \text{ as } r \rightarrow \infty,
\end{align} 
so that the solution in Eq.~\eqref{eq:VoPRadialEqnScattering} diverges at infinity unless there is cancellation between the two terms. A more detailed calculation confirms that total cancellation does not occur, and hence $\psi_{\lm0} \sim r$. Indeed, considering the dominant terms of Eq.~\eqref{eq:ScalarEOMFD} at large $r$, 
\begin{align}
	\frac{d^2\psi_{\lm0}}{dr^2} - \frac{\ell(\ell+1)}{r^2}\psi_{\lm0} = \frac{S_1}{r}\label{eq:leading_order_static_eqn}
\end{align}	
(where $S_1$ is a certain constant), we have the particular solution
\begin{align}
	\psi_{\lm0}(r) = - \frac{S_1 r}{\ell(\ell+1)}.
\end{align}
This is genuine behaviour; there is no homogeneous solution with matching large-$r$ behaviour that we can subtract off to remove the divergence.  For $\ell = 0$, a similar argument suggests that the behaviour is 
\begin{align}
	\psi_{000} \sim r\log r
\end{align}
at large $r$, which is what one would obtain by truncating the logarithmically divergent integral $c_{000}^-(r)$ at a finite upper integration limit in the variation of parameters formula, Eq.~\eqref{eq:VoPRadialEqnScattering}. 

In terms of the scalar field $\Phi$ itself [recall Eq.~\eqref{eq:spherharmPhi}], this behaviour translates to $\Phi^{\ell>0}\sim\text{const}$ and $\Phi^{\ell=0}\sim\log r$ as $r\rightarrow\infty$. Perhaps surprisingly, the static contributions to the scalar field do not fall off at infinity. Nonetheless Eq.~\eqref{eq:VoPRadialEqnScattering} gives the most regular solution for $\omega = 0$ and $\ell > 0$, in the sense that any other solution would either be irregular on the horizon or diverge as $r \rightarrow \infty$. For $\omega = 0 = \ell$, Eq.~\eqref{eq:VoPRadialEqnScattering} does not give the correct retarded solution (the integral in the second line is indefinite), but the true solution must diverge like $\psi_{000}(r) \sim r\log r$ as $r \rightarrow \infty$. This does not, however, mean that the time-domain solution diverges at infinity, and there is no sign of such behaviour in our numerical results. 

\section{Self-force from frequency modes}\label{sec:SF_from_modes}
In this section we will review the construction of the self-force from frequency modes of the scalar field, based on mode-sum regularization \cite{BarackOri2000}. We will discuss the problem of reconstructing the $\lm$-modes of the time-domain field, and review the method of EHS \cite{Barack:2008ms} conventionally used to achieve this, focusing on the challenge of modifying this approach to work with unbound orbits.

\subsection{Mode-sum regularization}\label{sec:mode_sum_regularization}
The self-force will be calculated using the standard mode-sum regularization procedure \cite{BarackOri2000, Barack2009}. The primary numerical inputs for this scheme are the ($m$-summed) $\ell$-mode contributions to the scalar field derivatives,
\begin{align}
    \SP{\nabla_{\alpha}\Phi}^{\ell}(x) :=  \sum_{m=-\ell}^{\ell}\nabla_{\alpha}\SB{\frac{1}{r} \psi_{\lm}(t,r)Y_{\lm}(\theta,\vf)}.
\end{align}
The scalar field and its derivatives are singular at the worldline, but the individual $\ell$-mode contributions are finite. The full force is defined by one-sided limits
\begin{align}
    F_{\alpha\pm}^{(\text{full})\ell}(\chi) = \lim_{x \rightarrow x_p^{\pm}(\chi)} q\SP{\nabla_{\alpha}\Phi}^{\ell},
\end{align}
where the $\pm$ denotes whether the limit is taken in the direction $r \rightarrow r_p(\chi)^+$ or $r \rightarrow r_p(\chi)^-$, and $\chi$ is being used as a parameter along the orbit.

The self-force may then be calculated using the mode-sum
\begin{align}
    F_{\alpha}^{\text{self}}(\chi) = \sum_{\ell=0}^{\infty}\SB{F_{\alpha\pm}^{(\text{full})\ell}(\chi) - \SP{\ell+\frac{1}{2}}A_{\alpha}^{\pm}(\chi) - B_{\alpha}(\chi)},\label{eq:mode_sum_formula}
\end{align}
where $A_{\alpha}^{\pm}$ and $B_{\alpha}$ are analytically known regularization parameters, which were first derived for generic Schwarzschild geodesic orbits in \cite{regpar, Barack:2001gx}. The terms in the mode-sum decay like $O(\ell^{-2})$. In practical calculations, the series is truncated at some $\ell = \lmax$, leaving a truncation error of $O(\lmax^{-1})$. 

This truncation error may be reduced by subtracting additional regularization-parameter terms in the mode sum \cite{DetMessWhit03}. These higher-order terms take the form
\begin{align}
    \frac{F^{[2]}_{\alpha}}{\SP{\ell-\frac{1}{2}}\SP{\ell+\frac{3}{2}}} + \frac{F^{[4]}_{\alpha}}{\SP{\ell-\frac{3}{2}}\SP{\ell-\frac{1}{2}}\SP{\ell+\frac{3}{2}}\SP{\ell+\frac{5}{2}} } +\cdots , 
\label{eq:higher_order_reg_terms}
\end{align}
where the coefficients $F^{[2n]}_{\alpha}$ are known analytically for generic Schwarzschild geodesic orbits for $n = 1, 2$ and $3$ \cite{Heffernan2012}. The additional terms in \eqref{eq:higher_order_reg_terms} sum to $0$ and thus do not change the value of the mode sum, but they do increase the rate of convergence. If the regularization terms up to and including $F^{[2n]}_{\alpha}$ have been subtracted, the terms in the mode sum decay like $O\SP{\ell^{-(2n+2)}}$, and the truncation error is $O\SP{\lmax^{-(2n+1)}}$.

\subsection{Conservative and dissipative pieces of the self-force}
When considering the physical effects of the self-force, it can be convenient to decompose it into ``conservative'' (time-symmetric) and ``dissipative'' (time-antisymmetric) pieces, 
\begin{align}
	F_{\alpha}^{\text{self}} = F_{\alpha}^{\text{cons}} + F_{\alpha}^{\text{diss}}\label{eq:cons_diss_split}
\end{align}
\cite{HindererFlanagan2008, Barack2009}, defined by
\begin{align}
	F_{\alpha}^{\text{cons}} &= \frac{1}{2}\left[F_{\alpha}^{\text{self(ret)}} + F_{\alpha}^{\text{self(adv)}}\right],\label{eq:cons_def}\\
	F_{\alpha}^{\text{diss}} &= \frac{1}{2}\left[F_{\alpha}^{\text{self(ret)}} - F_{\alpha}^{\text{self(adv)}}\right].\label{eq:diss_def}
\end{align}
Here $F_{\alpha}^{\text{self(ret)}}$ is the usual self-force constructed from the retarded scalar field, and $F_{\alpha}^{\text{self(adv)}}$ is the self-force constructed in just the same way, but from the scalar field obeying advanced boundary conditions. 

Taking advantage of the symmetries of Schwarzschild geodesics, it can be shown that the advanced self-force may be related to the retarded self-force by \cite{Mino:2003yg, HindererFlanagan2008} 
\begin{align}
 	F_{\alpha}^{\text{self(adv)}}(\chi) = \epsilon_{\alpha}F_{\alpha}^{\text{self(ret)}}(-\chi),\label{eq:adv_from_ret}
\end{align} 
where $\epsilon_{\alpha} = (-1, 1, 1, -1)$ in Schwarzschild coordinates, the periapsis is at $\chi = 0$ and there is no sum over $\alpha$ on the right-hand side. Equation~\eqref{eq:adv_from_ret} thus provides a practical means to calculate the conservative and dissipative pieces of the self-force using Eq.~\eqref{eq:cons_def} and \eqref{eq:diss_def}, without having to calculate the advanced field. 

The mode sums for the advanced and retarded self-force give mode sums for the conservative and dissipative pieces of the force  \cite{Barack2009},
\begin{align}
	F^{\text{cons}}_{\alpha} &= \displaystyle\sum_{\ell=0}^{\infty}\left[F^{\text{full(cons)}\pm}_{\alpha, \ell} - \SP{\ell+\frac{1}{2}}A_{\alpha}^{\pm}- B_{\alpha}\right] \label{eq:modesum_formula_cons},\\
	F^{\text{diss}}_{\alpha} &= \displaystyle\sum_{\ell=0}^{\infty}F^{\text{full(diss)}\pm}_{\alpha, \ell}\label{eq:modesum_formula_diss},
\end{align} 
where 
\begin{align}
	F^{\text{full(cons)}\pm}_{\alpha,\ell} &= \frac{1}{2}\left[F^{\text{full(ret)}\pm}_{\alpha,\ell} + F^{\text{full(adv)}\pm}_{\alpha,\ell}\right],\\
	F^{\text{full(diss)}\pm}_{\alpha,\ell} &= \frac{1}{2}\left[F^{\text{full(ret)}\pm}_{\alpha,\ell} - F^{\text{full(adv)}\pm}_{\alpha,\ell}\right].
\end{align}
In particular, we note that the terms in the mode-sum for $F^{\text{cons}}_{\alpha}$ require regularization and decay at the same rate as the total self-force, while $	F^{\text{diss}}_{\alpha}$ does not require regularization and the mode sum converges exponentially \cite{Barack2009}. The convergence of the mode sum for he conservative piece may be accelerated by subtracting higher-order regularization parameters. 

In Sec.~\ref{sec:scatter_angle_calc} we will consider the separate effects of the conservative and dissipative pieces of the self-force on the scatter angle.

\subsection{Method of extended homogeneous solutions}\label{sec:EHS_section}
The primary inputs for the mode-sum formula 
are the  $\lm$-modes of the time-domain scalar field and its derivatives at the particle. A naive attempt to obtain these from the frequency-domain field in Eq.~\eqref{eq:VoPRadialEqnScattering} faces the problem of the Gibbs phenomenon, first discussed in this context in Ref.~\cite{Barack:2008ms}. The presence of a Dirac delta function source, supported on the worldline, on the right hand side of Eq.~\eqref{eq:ScalarEOMTD} causes the derivatives $\psi_{\lm, t}$ and $\psi_{\lm, r}$ to be discontinuous at the worldline. A standard result from Fourier analysis says that the Fourier series/integrals for these derivatives will then converge to the correct value everywhere off the worldline, but will do so slowly [with terms decaying like $O\SP{\omega^{-1}}$] and non-uniformly in the vicinity of the particle. On the worldline, the series would be expected to converge to the 2-sided average of the derivative.  

The solution to this problem was developed in Ref.~\cite{Barack:2008ms} for bound particle motion. It involves expressing the field on either side of the worldline [i.e. in $r \leq r_p(t)$ and $r \geq r_p(t)$] in terms of analytic, homogeneous, frequency modes that have exponentially convergent Fourier series. Here we present the argument used in \cite{Barack:2008ms}, adapted to the unbound problem, to reconstruct the field in the interior region $r \leq r_p(t)$. 

First one defines the \textit{internal extended homogeneous solution},
\begin{align}
   \tilde\psi_{\lm\omega}^{-}(r) := C_{\lm\omega}^-\psi_{\lw}^-(r),
\end{align}
where the normalization integral $C_{\lm\omega}^-$ is as defined in Eq.~\eqref{eq:Cminus_def}. With this choice, the internal EHS is equal to the inhomogeneous field $\psi_{\lm\omega}$ when $ r \leq \rmin$. Defining the corresponding time-domain EHS field to be $\tilde\psi_{\lm}^{-}(t,r)$, we thus have 
\begin{align}
    \psi_{\lm}(t,r) = \tilde\psi_{\lm}^{-}(t,r) \quad \text{ for } r  \leq \rmin.
\end{align}
It was demonstrated in Ref.~\cite{Barack:2008ms} that $C_{\lm\omega}^-$ is expected to decay exponentially with $\omega$, such that $\tilde\psi_{\lm}^{-}(t,r)$ is analytic in both $t$ and $r$. The inhomogeneous field $\psi_{\lm}(t,r)$ is likewise expected to be analytic everywhere in $r < r_p(t)$, and agrees with $\tilde\psi_{\lm}^{-}(t,r)$ in the open set $r < \rmin$. We must therefore have an equality throughout the domain,
\begin{align}
     \psi_{\lm}(t,r) = \tilde\psi_{\lm}^{-}(t,r) \quad \text{ for } r  < r_p(t).
\end{align}

Crucial to this argument is the existence of the vacuum region $r \leq \rmin$ where the inhomogeneous field coincides with a homogeneous solution. In the case of bound motion, there is also a vacuum region $r \geq \rmax$ that allows the definition of an external EHS field that may be used to reconstruct the field in the region $r \geq r_p(t)$. Such an external vacuum region does not exist for a hyperbolic orbit, and the EHS method is constrained to calculating only the time-domain scalar field in the region $r \leq r_p(t)$.

The mode-sum regularization approach outlined in Sec.~\ref{sec:mode_sum_regularization} is usually implemented using a two-side average of the limiting values corresponding to $x\to x_p^\pm$, which simplifies the form of the regularization parameters. However, this is not strictly necessary, and it is possible to carry out the regularization procedure ``one-side'', making use of only the value of the scalar field derivatives taken from the direction $r < r_p(t)$. This is the approach we will be taking in this paper. 

We note here the small-frequency behaviour of  $\tilde\psi_{\lmw}^{-}(r)$, which will play a role later. For $r \ll \omega^{-1}$, the homogeneous solution $\psi_{\lw}^-(r)$ behaves like the polynomially growing static solution in Eq.~\eqref{eq:static_homsol_asymp}, growing approximately proportionally to $r^{\ell+1}$. We thus have that
\begin{align}
	\tilde\psi_{\lmw}^{-}(r_p) \sim \left(\frac{r_p}{r_{\rm min}}\right)^{\ell+1}\psi_{\lmw}(r_{\rm min}).\label{eq:EHS_growth}
\end{align}
for $\rmin \leq r_p \ll \omega^{-1}$, where we made use of the fact that the internal EHS and the physical, inhomogeneous, field coincide in $r \leq \rmin$. At small frequency the internal EHS field at the particle thus grows as a power law in $r_p$, and exponentially in $\ell$.  It was noted in Ref.~\cite{vandeMeent2016} that this behaviour results in significant cancellation between low-frequency modes when the EHS method is used to reconstruct the time-domain field away from the orbital turning points, with subsequent loss of precision. This problem and its implications will be further explored in Section \ref{sec:cancellation_problem}.

\section{Convergence of the radial integral}\label{sec:radial_integral}

When calculating the scalar-field self-force using the EHS method and one-sided regularization, a key numerical task is to evaluate the normalization integrals $C_{\lmw}^-$ defined in Eq.~\eqref{eq:Cminus_def}. As discussed in Section~\ref{sec:inhom_solution}, this integral displays marginal $\sim\text{oscillations}/r$ convergence. In this section we will explore the consequences of truncating the normalization integral at a finite radius, and develop two techniques to suppress the resulting truncation error. The \textit{tail correction scheme} will involve a series of analytical approximations to the neglected tail of the integral, while \textit{integration by parts} will be used to increase the decay rate of the original integrand. 

\subsection{Truncating the normalization integral}\label{sec:truncation_problem}
\begin{figure}
    \centering
  \includegraphics[width=0.49\textwidth]{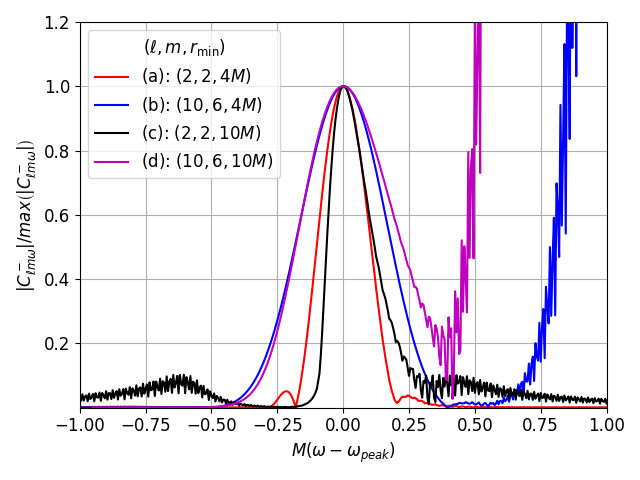}~\\~\\
  \caption[Illustration of the truncation problem]{\label{fig:truncation_problem}Example spectra of $|C_{\lmw}^-|$, shifted by the peak frequency and normalized to unity at the peak. The modes $(\ell,m)=(2,2)$ and $(10,6)$ are illustrated, for the geodesics with fixed $E = 1.1$ and $\rmin \in \{4M, 10M\}$. The peak frequencies were $M\omega_{\rm peak} \approx 0.29, 0.96, 0.095$ and $0.45$ for modes (a)--(d), respectively. The numerical integrals were truncated at $\rmax = 2000M$. High-frequency noise is visible in the tails of the spectrum as a result of this truncation. 
  } 
\end{figure}

In order to calculate the normalization integral $C_{\lmw}^-$ numerically, we wish to truncate it at some finite radius $\rmax$. Figure~\ref{fig:truncation_problem} displays some examples of $C_{\lmw}^-$ spectra for the two geodesics with $(E,\rmin) = (1.1,4M)$ and $(1.1,10M)$. These integrals were calculated numerically using the numerical methods discussed in Sec.~\ref{sec:numerical_method}, truncating the integral at $\rmax=2000M$. 
The modes $(\ell,m)=(2,2)$ and $(10,6)$ are displayed for both geodesics. We note that as we move away from the peak, high-frequency noise appears in the spectra and the numerical curves break away from the expected decaying trend. 
For a given energy and fixed $\rmax$, the problem is more acute for geodesics with larger $\rmin$ and at larger $\ell$. 

The high-frequency noise is not a numerical artefact, and can be traced to the truncation of $C_{\lmw}^-$ at finite radius. Because of the marginal oscillations/$r$-type convergence, suppressing this issue by increasing $\rmax$ is impractical. 
Instead, we introduce two analytical mitigation techniques, with negligible increase in computational cost. The first technique is based on an analytical approximation of the large-$r$ truncated tail of the $C_{\lmw}^-$ integral, and the second uses integration by parts to improve the convergence of the integral. In the rest of this section we discuss each technique in turn, and in Sec.\ \ref{sec:FD_results} we will illustrate their effectiveness in enabling a fast and efficient evaluation of $C_{\lmw}^-$.

\subsection{Tail correction scheme}\label{sec:correction_scheme}
We write $C_{\lmw}^-$, defined in Eq.~\eqref{eq:Cminus_def}, in the form 
\begin{equation}\label{eq:Cminus_Jlmw}
C_{\lmw}^- = -\frac{4qd_{\lm}}{W_{\lw}}\int_{r_{\rm min}}^\infty  J_{\lmw}(r)dr,
\end{equation}
where
\begin{align}
J_{\lmw}(r) &= \frac{1}{2}\displaystyle\sum_{\sigma=\pm 1}\frac{\psi_{\lw}^+(r)\exp{\left[i\sigma\left(\omega t_p(r)- m\varphi_p(r)\right)\right]}}{r \left|\dot{r}_p(r)\right|}.\label{eq:integrandminus}
\end{align}
We seek to obtain a large-$r$ asymptotic expansion for $J_{\lmw}$. Starting with the  homogeneous solution factor, we 
 recall from Eq.\ (\ref{eq:asymp_inf}) its asymptotic form, 
\begin{align}
	\psi_{\lw}^+(x) &= e^{i\tilde\omega x_*}\sum_{k = 0}^{\infty} c_k^{\infty} x^{-k} ,\label{eq:asymp_inf_tilder}
\end{align}
where for convenience we have introduced here $x:=r/(2M)$, as well as $x_*:=r_*/(2M)$ and $\tilde\omega:=2M\omega$.
Using the identity 
\begin{align}
	x_* = x + \log x - \sum_{n = 1}^{\infty}\frac{1}{n}\, x^{-n},
\end{align}
this becomes
\begin{align}
	\psi_{\lw}^+(x) = e^{i\tilde{\omega}x}x^{i\tilde{\omega}}\left(1 + \sum_{n =1}^\infty\hat{c}_n^{\infty}x^{-n}\right), \label{c_hat}
\end{align}
where the new expansion coefficients, $\hat{c}_n^{\infty}$, can be written in terms of the old ones, $c_n^{\infty}$. In Appendix~\ref{app:correction_scheme_constants} we give the explicit relations for $1\leq n \leq 5$, which will suffice for our purpose. 

We turn next to the phase factor in \eqref{eq:integrandminus}. Using Eqs.~\eqref{eq:texp} and \eqref{eq:phiexp}, we obtain
\begin{align}
	\omega t_p - m\varphi_p = \tilde{\omega}\left(\frac{x}{v} + B\log x\right) + \displaystyle\sum_{n\geq 0} \Delta_n x^{-n},\label{eq:phase_expansion}
\end{align}
where 
\begin{align}
\Delta_0 := \omega t_0 - m\varphi_{\infty}\quad\text{and}\quad \Delta_{n > 0} := \tilde{\omega}C_n - mD_n, \label{Delta_inf}
\end{align}
with
$B, t_0, \varphi_{\infty}, C_n$ and $ D_n$ being the coefficients appearing in the expansions \eqref{eq:texp} and \eqref{eq:phiexp}. We further recall the large-$r$ expansion of $1/|\dot{r}_p|$ from Eq.~\eqref{eq:urinv_exp},
\begin{align}
	\frac{1}{|\dot{r}_p|} = \sum_{n=0}^{\infty}U_nx^{-n}.
\end{align}
The complex exponential function is then expanded as
\begin{align}
	\exp\left[i\sigma\sum_{n\geq 0}\Delta_n x^{-n}\right] = e^{i\Delta_0}\left[1+ \sum_{n\geq 1}H_{n\sigma}x^{-n}\right]\label{eq:corr_exponential_expansion},
\end{align}
where expressions for $H_{n\sigma}$ are given in terms of $\Delta_n$ for $n \leq 5$ in Appendix~\ref{app:correction_scheme_constants}. 

Substituting the above expansions, Eq.~\eqref{eq:integrandminus} takes the form
\begin{align}\label{eq:integrandexpansion}
	&J_{\lmw}(x) = \frac{1}{2}\displaystyle\sum_{\sigma = \pm1}\displaystyle\sum_{n\geq0}\lambda_{n\sigma}e^{i\tilde{\Omega}_{\sigma}x}x^{a_{n\sigma}-1},
\end{align}
as $x \rightarrow \infty$, where 
\begin{align}
    \tilde{\Omega}_{\sigma} &:= (1+\sigma/v)\tilde{\omega} \label{eq:tOmega_def}, \\a_{n\sigma} &:= i(1+\sigma B)\tilde{\omega} - n,\label{eq:a_def}
\end{align}
and
\begin{align}
	\lambda_{n\sigma} := \frac{1}{2M}e^{i\sigma\Delta_0}
\displaystyle\sum_{q+r+s=n}U_q\hat{c}_r^{\infty}H_{s\sigma},
\end{align}
where the sum is taken over non-negative integers $q, r, s$, and we define $H_{0\sigma} = 1 = \hat{c}^{\infty}_0$. 

A key observation is that the expression in \eqref{eq:integrandexpansion} can be integrated {\it analytically}, term by term:
\begin{align}
	\displaystyle\int_{x_{\rm max}}^{+\infty}&J_{\lmw}(x')dx' \nonumber \\ &= \frac{1}{2}\displaystyle\sum_{\sigma = \pm 1}\sum_{n\geq 0}\lambda_{n\sigma}\SP{-i\tilde{\Omega}_{\sigma}}^{-a_{n\sigma}}\Gamma[a_{n\sigma}, z_{\sigma}], \label{eq:higher_order_correction}
\end{align}
where $x_{\rm max} = r_{\rm max}/2M$, $a_{n\sigma}$ depends on $\sigma$ and $n$ through Eq.~\eqref{eq:a_def}, $z_{\sigma} := -i\tilde{\Omega}_{\sigma}x_{\rm max}$, and $\Gamma[a, z]$ is the upper incomplete gamma function, calculated in practice using the continued fraction representation,
\begin{align}
    \Gamma[a,z] = \cfrac{z^a e^{-a}}{z + \cfrac{1-a}{1+ \cfrac{1}{z+ \cfrac{2-a}{1+ \cfrac {2}{z + \ddots} }  } } }.
\end{align}
 As $x_{\rm max}\rightarrow\infty$, the $n$-th term of the series in Eq.~\eqref{eq:higher_order_correction} has the asymptotic behaviour
\begin{align}
    \left|\Gamma[a_{n\sigma}, z_{\sigma}]\right| \sim \frac{1}{x_{\rm max}^{n+1}}.\label{eq:gamma_asymptotics}
\end{align}

The tail expression \eqref{eq:higher_order_correction} may be added to a numerical integral that has been truncated at finite radius $r_{\rm max}$, to obtain a more accurate estimate of $C_{\lmw}^-$. We define a \textit{tail correction scheme of order $N$} as one obtained by truncating Eq.~\eqref{eq:higher_order_correction} at finite order $n = N - 1$, and using this expression as an approximation to the tail of the integral $C_{\lmw}^-$. In this scheme, terms up to and including $r^{-N}$ are included in the expansion of the integrand, and the error in the tail estimate is $O\left(r_{\rm max}^{-(N+1)}\right)$, where $r_{\rm max}$ is the radius at which the numerical portion of the integral is truncated. This error estimate is obtained by applying Eq.~\eqref{eq:gamma_asymptotics} to the first neglected term, $n = N$, of Eq.~\eqref{eq:higher_order_correction}. For a correction scheme of order $N$, one requires the coefficients $\lambda_{n\sigma}$ up to and including $n = N - 1$. We have obtained these coefficients up to $n=5$, sufficient to implement the tail correction scheme at orders up to $N=6$. 

\subsection{Integration by parts}\label{sec:IBP}
The key to the integration by parts (IBP) approach is to factorize the integrand $J_{\lmw}$ in Eq.\ (\ref{eq:Cminus_Jlmw}) into (a) a sinusoidal factor that may be integrated repeatedly, and (b) a decaying factor that may be practically differentiated without recourse to numerical differentiation, and which decays more rapidly each time it is differentiated. In this section, we will demonstrate the existence of such a factorization, and thus show how IBP may be used to increase the rate of convergence of the integrals $C_{\lmw}^-$. 

First, however, we make a remark about the behaviour of the integrand $J_{lm\omega}$ at the lower boundary, $r=\rmin$. As can be seen from Eq.~\eqref{eq:integrandminus}, there is a factor of $|\dot{r}_p|\propto (r-\rmin)^{1/2}$ present in the denominator, which results in an integrable singularity. Although this does not prevent the integral converging, it is numerically problematic, and confounds an attempt at integration by parts, which may introduce a stronger, non-integrable singularity. 

To handle this issue, we select some radius $\rcut > \rmin$, and use the relativistic anomaly $\chi$ as the integration variable in the region $\rmin\leq r \leq \rcut$. The use of $\chi$ as integration variable produces an integrand which is regular at $\rmin$, but which suffers from increasingly rapid, large amplitude oscillations as $\chi\rightarrow\chiinf$. It is therefore more practical to use $r$ as the integration variable for the $r > \rcut$ leg of the integral. A similar approach was adopted in Ref.~\cite{Hopper2018}. The practical details of the integration, including the choice of $\rcut$, will be discussed in Section~\ref{sec:numerical_method}.

We thus wish to apply IBP to the integral
\begin{equation}\label{eq:Cminus(r)}
C_{\lmw}^{(r)-} := \int_{r_{\rm cut}}^\infty  J_{\lmw}(r')dr,
\end{equation}
where the integrand $J_{\lmw}$, recall, is given in Eq.\ \eqref{eq:integrandminus}.
To obtain the sinusoidal factor we desire, we consider the phases of the oscillatory factors in the integrand, and remove the parts that grow linearly with $r$ at large radii. Define
\begin{align}
	\Delta(r) := \omega t_p(r) - m \varphi_p(r) - \frac{\omega r}{v},\label{eq:Delta_def}
\end{align}
and note that Eq.~\eqref{eq:phase_expansion} implies
\begin{align}
	\Delta(r) = \wbar B\log\left(x\right) + \sum_{n\geq 0}\Delta_nx^{-n}\label{eq:delta_expansion_IBP}
\end{align}
as $r \rightarrow\infty$, where the coefficients $\Delta_n$ are the same as those defined in Sec.~\ref{sec:correction_scheme} for the correction scheme. Equation~\eqref{eq:delta_expansion_IBP} then implies that
\begin{align}
	\frac{d}{dr}\Delta(r) = O\left(r^{-1}\right)
\end{align}
as $r\rightarrow \infty$. We rewrite Eq.~\eqref{eq:Delta_def} as
\begin{align}
	\Delta(r) = \omega\hat{t}_p(r) - m\varphi_p(r),
\end{align}
introducing the new time coordinate 
\begin{align}
	\hat{t}_p(r) := t_p(r) - r/v,
\end{align}
which diverges logarithmically as $r\rightarrow \infty$. $\hat{t}_p$ can be calculated directly by integrating, 
\begin{align}
	\frac{d\hat{t}_p}{d\chi} =\> &\frac{M}{(1+e\cos\chi)^2}\Bigg[-\frac{pe\sin\chi}{v} \label{eq:dthatdchi} \\ \nonumber&+\frac{p^2}{p-2-2e\cos\chi}\sqrt{\frac{(p-2)^2-4e^2}{p-6-2e\cos\chi}} \Bigg],
\end{align} 
obtained using Eqs.~\eqref{eq:dt_dchi} and \eqref{eq:ep_def}.  
We also introduce the new field 
\begin{align}
	P_{\lw}(r) := e^{-iwr}\psi_{\lw}^+(r),\label{eq:P_def}
\end{align}
which, using \eqref{eq:ScalarEOMFD_hom}, satisfies
\begin{align}
	\frac{d^2P_{\lw}}{dr_*^2} + &2i\omega f(r) \frac{dP_{\lw}}{dr_*} + \label{eq:fieldeqn_P}\\ \nonumber &\left[-V_l + \omega^2 - \omega^2 f(r)^2 + \frac{2iM\omega}{r^2}f(r)\right]P_{\lw}(r)=0.
\end{align}
The large-$r$ behaviour of $P_{\lw}$ is easily deduced from that of $\psi_{\lw}^+$ given in Eq.~\eqref{eq:asymp_inf}:
\begin{align}
	P_{\lw} =\exp\left[2iM\omega\log\left(\frac{r}{2M}-1\right)\right]\sum_{k = 0}^{\infty} c_k^{\infty}\left(\frac{2M}{r}\right)^k \label{eq:asymp_inf_P}
\end{align}
as $r\rightarrow\infty$. Note $P_{\lw}$ oscillates with only logarithmic phase at infinity. 

With these definitions of $\Delta(r)$ and $P_{\lw}(r)$, we obtain
\begin{align}
	C_{\lmw}^{(r)-} &= \frac{1}{2}\sum_{\sigma=\pm 1}\displaystyle\int_{\rcut}^{+\infty}e^{i\Omega_{\sigma}r} K_{\lmw}^{\sigma}(r)dr,\label{eq:IBP_integrand_form}
\end{align}
where
\begin{align}
    \Omega_{\sigma} := (1+\sigma/v)\omega
\end{align}
is related to Eq.~\eqref{eq:tOmega_def} by $\Omega_{\sigma} = \tilde{\Omega}_{\sigma}/2M$, and
\begin{align}
	K_{\lmw}^{\sigma}(r) := \frac{P_{\lw}(r)e^{i\sigma\Delta(r)}}{r|\dot{r}_p(r)|}.\label{eq:IBP_K_def}
\end{align}
We note two properties of the function $K_{\lmw}^{\sigma}$. First, we have closed form expressions for $\dot{r}_p(r)$ using Eq.~\eqref{eq:RadialEqn},
and also
\begin{align}
	\frac{d}{dr}\Delta(r) = \frac{\omega E}{f(r)\dot{r}_p(r)} - \frac{mL}{r^2 \dot{r}_p(r)} - \frac{\omega}{v},
\end{align}
both of which can be differentiated in closed analytical form any number of times. We can also determine $P_{\lw}$ and $dP_{\lw}/dr$ numerically, and then recursively determine any number of derivatives using the field equation \eqref{eq:fieldeqn_P}. Thus we may practically differentiate $K_{\lmw}^{\sigma}$ any given number of times using repeated applications of the product rule.

Second, each derivative of $K_{\lmw}^{\sigma}$ decays one order more rapidly in $r$ than the previous derivative:
\begin{align}
    K_{\lmw}^{\sigma(N)} = O\left(\frac{1}{r^{N+1}}\right)\label{eq:Kderiv_falloff}
\end{align}
as $r \rightarrow\infty$, where we introduced the derivative notation $K_{\lmw}^{\sigma(N)} := d^NK_{\lmw}^{\sigma}/dr^N$. To show this, it suffices to show that each factor in Eq.~\eqref{eq:IBP_K_def} decays one order more rapidly each time it is differentiated. This is easily confirmed using the closed-form expressions for the derivatives of $e^{i\sigma\Delta}$ and of $1/(r|\dot{r_p}|)$. The result for the $P_{\lw}$ factor follows from Eq.~\eqref{eq:asymp_inf_P}. Crucial to this result was the fact that the two oscillatory factors $e^{i\sigma\Delta}$ and $P_{\lw}$ oscillate with only logarithmic phase as $r \rightarrow \infty$ and hence have decaying derivatives. 

Equation~\eqref{eq:IBP_integrand_form} thus provides the desired factorization of the integrand. Integrating by parts $N+1$ times gives
\begin{align}
	C_{\lmw}^{(r)-} =  \frac{1}{2}\displaystyle\sum_{\sigma=\pm 1}\left\{\sum_{n=0}^N \left[\left(\frac{i}{\Omega_{\sigma}}\right)^{n+1}e^{i\Omega_{\sigma}r_{cut}}K_{\lm\omega}^{\sigma(n)}(r_{cut})\right] \right.
 \nonumber\\
 +  \left.\left(\frac{i}{\Omega_{\sigma}}\right)^{N+1}\displaystyle\int_{r_{cut}}^{+\infty}e^{i\Omega_{\sigma}r} K_{\lm\omega}^{\sigma(N+1)}(r)dr\right\} \label{eq:IBPformula}.
\end{align}
We can practically apply integration by parts any number of times, and hence achieve any polynomial rate of decay in the integrand. Using Eq.~\eqref{eq:Kderiv_falloff}, we see that the integrand in Eq.~\eqref{eq:IBPformula} decays like $r^{-(N+2)}$ as $r \rightarrow\infty$. The limiting factor of the IBP method is the need to derive expressions for the necessary derivatives of $K_{\lm\omega}^{\sigma}$ in advance, which becomes increasingly complicated at high orders. In practice we have only derived the expressions for derivatives up to and including $K_{\lm\omega}^{\sigma(4)}$, allowing for four iterations of IBP and a truncation error of $O(r_{\rm max}^{-5})$.

 It is possible to derive a tail correction scheme, analagous to Eq.~\eqref{eq:higher_order_correction},  to approximate the tail of the integral in Eq.~\eqref{eq:IBPformula}. Indeed, comparing the form of the integrand in Eq.~\eqref{eq:IBP_integrand_form} to the expansion in Eq.~\eqref{eq:integrandexpansion}, we can read off the series expansion for $K_{\lmw}^{\sigma}$ at large $r$:
\begin{align}
	K_{\lmw}^{\sigma} = \sum_{n \geq 0} \lambda_{n\sigma}x^{a_{n\sigma}- 1},\label{eq:Kexp}
\end{align}
where $x = r/(2M)$ as usual. Differentiating this term by term, we have
\begin{align}
	\left(\frac{i}{\Omega_{\sigma}}\right)^pK_{\lmw}^{\sigma(p)} = \sum_{n \geq 0} \lambda_{np\sigma}x^{a_{n+p,\sigma}-1},\label{eq:derivKexp}
\end{align}
where the new coefficients $\lambda_{np\sigma}$ are given by
\begin{align}
	\lambda_{np\sigma} = \left(\frac{i}{\tilde{\Omega}_{\sigma}}\right)^p\lambda_{n\sigma}\prod_{q=0}^{p-1}\left[a_{n+q,\sigma}-1\right].\label{eq:lambda_tilde_def}
\end{align}
The tail of the IBP integral may then be approximated using
\begin{align}
	&\left(\frac{i}{\Omega_{\sigma}}\right)^p\displaystyle\int_{x_{\rm max}}^{+\infty}dx\>e^{i\Omega_{\sigma}r} K_{\lmw}^{\sigma(p)} \nonumber \\
	&\approx \frac{1}{2}	\sum_{n=0}^{n_{\rm max}}\displaystyle\int_{x_{\rm max}}^{+\infty}dx\>\lambda_{np\sigma}e^{i\tilde{\Omega}_{\sigma}x}x^{a_{n+p,\sigma}-1} \nonumber \\
	& = \frac{1}{2}\sum_{n=0}^{n_{\rm max}}\lambda_{np\sigma}\SP{-i\tilde{\Omega}_{\sigma}}^{-a_{n+p,\sigma}} \Gamma[a_{n+p,\sigma},z_{\sigma}], 
\end{align}
where, $a_{n\sigma}$ is as defined in Eq.~\eqref{eq:a_def} and again $z_{\sigma} = -i\tilde{\Omega}_{\sigma}x_{\rm max}$.


\section{Numerical method}\label{sec:numerical_method}
In this section we present the details of our numerical approach, implemented in C.  We start with the numerical calculation of the homogeneous solutions $\psi_{\lw}^{\pm}$, and then describe the quadrature routines used to calculate the normalization integrals $C_{\lmw}^-$, and the details of our application of the tail correction and IBP methods. We discuss the numerical approach taken to evaluate the inverse Fourier integrals needed to obtain the time-domain $\ell$-modes of the scalar field derivatives, which we use as input to the mode-sum regularization scheme. In particular we highlight how this can be done efficiently by interpolating the normalization integrals over frequency, minimizing the number of expensive integral evaluations.

\subsection{Homogeneous solutions}\label{sec:method_homsols}

Boundary conditions for the homogeneous solution $\psi_{\lw}^-$ and its radial derivative are provided by the series in Eq.~\eqref{eq:asymp_eh}. The coefficients $c_{k>0}^{\rm eh}$ are obtained by recursively using the relation in Appendix~\ref{app:homsol_BCs}, with initial conditions $c_{k<0}^{\rm eh}=0$ and $c_0^{\rm eh}=1$. Successive terms in the series are calculated and added, stopping when the relative contribution of the last term falls below some threshold, usually taken to be $10^{-16}$. The boundary conditions for $\psi_{\lw}^-$ are specified at radius $r_{*}^{\rm in} = -60M$. This value is limited by machine precision when inverting the relation $r_*(r)$ to get $r(r_*)$; for $r_*^{\rm in} < -60M$, $r$ begins to become indistinguishable from $2M$ at double precision. Despite this, we find that this choice of $r_*^{\rm in}$ is adequate for rapid convergence of series~\eqref{eq:asymp_eh}.

The field $\psi_{\lw}^-$ is then obtained at radii $r_* > r_*^{\rm in}$ by evolving the initial data according to the homogeneous equation~\eqref{eq:ScalarEOMFD_hom}. This is done numerically using the Runge-Kutta Prince-Dormand (8,9) method rk8pd implemented in the GNU Scientific Library (GSL) \cite{GSL_lib}, with a requested relative error tolerance of $10^{-12}$. 

The calculation of the field $P_{\lw} = e^{-i\omega r}\psi_{\lw}^+$ is similar. Boundary conditions are obtained from the series in Eq.~\eqref{eq:asymp_inf_P} and its derivative, evaluated at some outer radius $r_*^{\rm out}$. The coefficients $c_{k>0}^{\infty}$ are once again obtained recursively using the relation described in Appendix~\ref{app:homsol_BCs} with initial conditions $c_{k<0}^{\infty}=0$ and $c_0^{\infty}=1$. The series is once again truncated when the relative contribution of the last term falls below $10^{-16}$. As noted in Ref.~\cite{Warburton:2011hp}, however, the terms in series \eqref{eq:asymp_inf_P} can begin to increase again after initially decreasing. This lack of convergence is unsurprising; the expansion in Eq.~\eqref{eq:asymp_inf_P} is only expected to converge in the wave zone $\omega r \gg 1$. For small frequency, the wave zone may lie beyond $r_*^{\rm out}$, in which case the series is not expected to converge. To resolve this we adopt a similar approach to Ref.~\cite{Warburton:2011hp}, increasing $r_*^{\rm out}$ in steps of $2000M$ until convergence is achieved. For an initial value of $r_*^{\rm out}$ we usually choose a value slightly larger than $r_*(\rmax)$, where $\rmax$ is the desired truncation radius for the normalization integral $C_{\lmw}^-$. 

Once the boundary conditions for $P_{\lw}$ have been calculated, the field at radii $r_* < r_*^{\rm out}$ is obtained by integrating Eq.~\eqref{eq:fieldeqn_P} inwards with respect to $r_*$. Once again we use the rk8pd routine from the GSL, with a relative error tolerance of $10^{-12}$. 

\subsection{Normalization integrals}\label{sec:method_Cminus}
As discussed briefly in Section~\ref{sec:IBP}, the radial integration is divided in two. In the region $\rmin \leq r < \rcut$ we use $\chi$ as the integration variable,
\begin{align}
    C_{\lmw}^{(\chi)-} := \displaystyle\int_{\chi=0}^{\chi(\rcut)} J_{\lmw}(r_p(\chi))\frac{dr_p}{d\chi}d\chi.
    \label{eq:Cminus(chi)}
\end{align}
The section over $\rcut \leq r \leq \infty$, which we earlier named $C_{\lmw}^{(r)-}$, then uses $r$ as the integration variable. $C_{\lmw}^{(r)-}$ is calculated from  Eq.~\eqref{eq:IBPformula} with the desired level of IBP applied; note that the integral is only to be truncated at finite radius $\rmax$ \textit{after} integration by parts has been applied. 

We make use of two quadrature routines from the GSL \cite{GSL_lib}. The first is the QAG general purpose adaptive integrator, which can be made to use Gauss-Kronrod rules with varying numbers of points. In particular, we make extensive use of QAG with the 61pt Gauss-Kronrod rule, an approach we refer to as QAG61. The second method we used is based on the QAWO routine, an adaptive integrator based around a 25pt Clenshaw-Curtis rule tailored towards integrands with a sinusoidal weight function. Such a routine is well suited for calculating $C_{\lmw}^{(r)-}$ because, as we have seen in Eqs.~\eqref{eq:IBP_integrand_form} and \eqref{eq:IBPformula}, the integrand can be factored into a sinusoidal factor and a factor that oscillates with only logarithmic phase. 

When evaluating $J_{\lmw}(r_p(\chi))$ in the integrand of $C_{\lmw}^{(\chi)-}$, we require the geodesic functions $t_p(\chi)$ and $\varphi_p(\chi)$. These are calculated by numerically integrating Eqs.~\eqref{eq:dt_dchi} and \eqref{eq:dphidchi} using the QAG61 routine with a relative error tolerance of $10^{-12}$. Evaluating $C_{\lmw}^{(r)-}$ additionally requires the modified time coordinate $\hat{t}_p = t_p - r/v$, which we calculate directly by integrating Eq.~\eqref{eq:dthatdchi} numerically using QAG61 with error tolerance $10^{-12}$. 
 
For a given $\ell$, $m$ and $\omega$, the approach taken to evaluate the normalization integrals $C_{\lmw}^-$ with a given truncation radius $\rmax$ is then as follows:
\begin{enumerate}
    \item The homogeneous solutions $\psi_{\lw}^-$ and $P_{\lw}$ are calculated using the method described in Section~\ref{sec:method_homsols} and stored at a dense sample of points in an interval containing $[\rmin, \rmax]$. The Wronskian $W_{\lw} = W[\psi_{\lw}^-, \psi_{\lw}^+]$ is also calculated. When, in subsequent steps, the homogeneous solutions are required at arbitrary radii $\rmin \leq r \leq \rmax$, these gridpoints are used as initial data and the homogeneous solution at the desired radius is obtained by integrating the appropriate field equation.  
    \item An initial choice of $\rcut = 2\rmin$ is selected. The integral $C_{\lmw}^{(\chi)-}$ is calculated numerically using the QAG61 routine. The relative error threshold is set at $10^{-10}$.
    \item If the integrator reports that it cannot achieve the error tolerance, $\rcut$ is reduced to $1.5\rmin$ and step (2) is repeated. This estimate is then stored, whether or not the new relative error estimate is less than $10^{-10}$.
    \item The integral $C_{\lmw}^{(r)-}$ is integrated by parts $X$ times before being truncated at $\rmax$, and then evaluated numerically with a relative error threshold of $10^{-10}$. This may be achieved using either the QAG61 or QAWO routine. 
    \item An order $Y$ correction to the neglected tail (appropriate to the number of iterations of IBP) is then added. 
\end{enumerate}

The above algorithm involves two parameters, $X$ and $Y$, which control the number of iterations of IBP and the order of the tail correction, respectively. We refer to such a method as \textit{IBPXcorrY}. For example, IBP0corr0 uses neither integration by parts nor adds any approximation for the neglected tail, while IBP4corr5 uses 4 iterations of IBP and a 5th order correction. 

\begin{figure}
  \centering
  \includegraphics[width=0.49\textwidth]{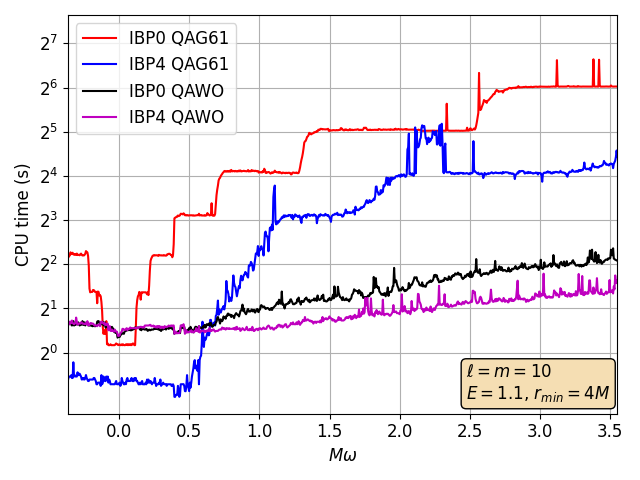}~\\~\\
  \caption[Timings of $C_{\lmw}^-$ using different methods.]{\label{fig:Cminus_timings}CPU time to calculate a single $C_{\lmw}^-$ integral (truncated at $\rmax=2000M$) as a function of frequency using different methods. The mode $\ell = m =10$ was selected for the geodesic with parameters $E = 1.1$, $\rmin = 4M$, and the test was carried out on a laptop computer with Intel i7-11850H processor (8 cores at 2.5 GHz). Different orders of IBP ($0$ vs $4$) and quadrature routines (QAG61 vs QAWO) for computing the $r\geq \rcut$ portion of $C_{\lmw}^-$ were tested. Using IBP has a modest time benefit in most circumstances, but the optimum quadrature routine depends on frequency.} 
\end{figure}

Figure~\ref{fig:Cminus_timings} displays the time taken to calculate $C_{10,10,\omega}^-$ as a function of $\omega$ for the geodesic with $E = 1.1$ and $\rmin=4M$, using different choices of IBP order and quadrature routine to evaluate $C_{\lmw}^{(r)-}$. At small frequencies, a single integral takes $\sim 1s$ when using QAG61 quadrature without any integration by parts, but this quickly rises in a stepwise fashion (almost doubling each step) and a single integral can exceed $60s$ at $|M\omega|\gtrsim 2.5$. Introducing IBP4 but maintaining QAG61 quadrature decreases the runtime to sub-$1s$ at small frequency, but the runtime still increases rapidly and may exceed $15s$ at the highest frequencies, a reduction of approximately $75\%$ compared to IBP0. QAWO quadrature produces a significant reduction in runtime at high frequencies, keeping runtimes below approximately $5s$ in this test, but at lower frequencies the QAG61 routine is faster. This is not unexpected; when the (sub-)interval length falls below a few wavelengths, the QAWO routine defaults to a 15pt Gauss-Kronrod rule, which is lower order than the 61pt rule we use with the QAG61 routine \cite{GSL_lib}. Integration by parts produces an approximately $40\%$ ($\sim 1.5s$) time saving when using QAWO quadrature at high frequencies, but makes little difference at low frequency.

For convenience we wish to adopt a single quadrature routine to evaluate $C_{\lmw}^{(r)-}$, to be used at all frequencies. It is therefore sensible to make use of the QAWO routine, because this is the faster routine at the majority of frequencies we require, and because this routine makes the largest absolute time savings. From here on, $C_{\lmw}^{(r)-}$ is always evaluated using the QAWO routine unless otherwise stated. 

One issue presents itself when attempting to use IBP at small frequencies. At small frequency, both the surface term and integral in Eq.~\eqref{eq:IBPformula} can grow very large, and there is a significant degree of cancellation between them. This results in a loss of precision, which, for some $\lm$ modes, creates a noisy spike in the $C_{\lmw}^-$ spectrum at small frequency. Fortunately a simple solution is available to this problem. By introducing an additional breakpoint $\rsplit$ between $\rcut$ and $\rmax$, one can evaluate the integral without IBP (or using low order IBP) in the interval $\rcut < r < \rsplit$, and then use a higher order of IBP for $r > \rsplit$. When $\rsplit \gg \rcut$, the error from cancellation between the surface term at $\rsplit$ and the integral over $r > \rsplit$ is much reduced. 

We refer to such split-order IBP methods as \textit{IBPXYcorrZ}, where $X$ and $Y$ are the orders of the IBP used in the regions $\rcut \leq r < \rsplit$ and $r > \rsplit$ respectively, and $Z$ is the order of the tail correction applied. In practice we only make use of IBP04corrZ methods in this paper. 

\subsection{Efficient time-domain reconstruction} \label{sec:method_TD_reconstruction}

In order to calculate the $\ell$-mode contributions to the $t$, $r$ and $\varphi$ derivatives of the scalar-field at a point $x_p^{\alpha}$ along the orbit, we have to numerically evaluate the inverse Fourier integrals
\begin{align}
	\Phi^{\lm-}_t&:= -\frac{1}{r_p}\displaystyle\int_{-\infty}^{+\infty}d\omega\> i\omega\>C_{\lmw}^-\psi_{\lw}^-(r_p)e^{-i\omega t_p},\label{eq:FullF_t}\\
	\Phi^{\lm-}_r &:= \displaystyle\int_{-\infty}^{+\infty}d\omega\> C_{\lmw}^-\partial_r\left(\frac{\psi_{\lw}^-(r)}{r}\right)_{r=r_p}e^{-i\omega t_p},\label{eq:FullF_r}\\
	\Phi^{\lm-}_{\varphi} &:= \frac{1}{r_p}\displaystyle\int_{-\infty}^{+\infty}d\omega\> im\>C_{\lmw}^-\psi_{\lw}^-(r_p)e^{-i\omega t_p},\label{eq:FullF_phi}
\end{align}
for $m \geq 0$ and $\ell+m$ even. Using the symmetry relation in Eq.~\eqref{eq:negmreln}, the $\ell$-modes of the full force are then given by
\begin{align}
	F^{(\text{full})\ell}_{\alpha-}&= q\Phi^{\ell0-}_{\alpha}Y_{\ell0}\left(\frac{\pi}{2}, \varphi_p\right) \nonumber \\ &\quad\quad\quad + 2q\sum_{m>0}\text{Re}\left[\Phi^{\lm-}_{\alpha}Y_{\lm}\left(\frac{\pi}{2}, \varphi_p\right)\right]\label{eq:SSF_sum_over_m},
\end{align}

for $\alpha = t, r, \varphi$, where only modes with $\ell + m = \text{even}$ contribute.

The most obvious way to evaluate integrals \eqref{eq:FullF_t}-\eqref{eq:FullF_phi} is to use an adaptive integrator such as the QAG61 routine we have made extensive use of. One issue with this approach is that an adaptive integrator will generically call different frequencies when evaluating different components, or when evaluating the integrals at different orbital positions. The oscillatory factors $\psi_{\lw}^-$ and $e^{-i\omega t_p}$ in the integrand may also require a denser sampling to resolve, particularly when $|t_p|$ is large, resulting in wasteful over-sampling of the normalization integrals. Given that a single $C_{\lmw}^-$ integral takes several seconds to compute in general (see Sec.~\ref{sec:method_Cminus}), it would be a very lengthy process to calculate $C_{\lmw}^-$ on the fly at every required frequency.

Interpolation provides one solution to this problem. In this approach, one first calculates the integrals $C_{\lmw}^-$ at a dense sample of frequency nodes $\omega_n$ for the $\lm$-modes required. The value of $C_{\lmw}^-$ at an intermediate frequency $\omega$ can then be estimated using interpolation. We do this by identifying the node $\omega_{N}$ that lies nearest to $\omega$, and then using the $2d$-degree polynomial that fits the data at the nodes $\omega_{N-d}$, $\omega_{N-d+1},...,\omega_{N+d}$. This interpolation is carried out in practice using the gsl\_interp\_polynomial interpolator type included in the GSL \cite{GSL_lib} and a degree 8 polynomial. 

Given a repository of $C_{\lmw}^-$ data at an appropriately dense sample of frequencies, integrals \eqref{eq:FullF_t}-\eqref{eq:FullF_phi} are then evaluated using the QAG61 routine with a relative error threshold of $10^{-8}$. These are summed over $m$ to get $F_{\alpha}^{(\text{full})\ell-}$ using Eq.~\eqref{eq:SSF_sum_over_m}.

\subsection{Overall approach}
In Sections~\ref{sec:method_homsols}-\ref{sec:method_TD_reconstruction} we discussed the methods used to calculate the homogeneous solutions and normalization integrals, and then how to efficiently calculate the $\lm$-mode contributions to the scalar field in the time domain, and hence the $\ell$ modes of the full force. We now outline how these blocks are combined to produce a calculation of the self-force along an orbit with parameters $(E, \rmin)$.

\textit{Step 1:} We find the maximal frequency limits that may be used before high-frequency noise appears in the spectra of $C_{\lmw}^-$. For a given $\lm$ mode, we begin calculating $C_{\lmw}^-$ at $\omega = m\omega_{\rm circ}$, corresponding to the frequency of a circular geodesic of radius $r_{\rm min}$, namely $M\omega_{\rm circ} = \left(M/\rmin\right)^{3/2}$. This frequency was used as an estimate of the peak frequency, and worked well in our tests. The frequency was then increased in steps $M\Delta\omega = 5\times 10^{-3}$, and at each step the following procedure was applied:
\begin{enumerate}
	\item If fewer than $6$ data points $(\omega, C_{\lmw}^-)$ are available, continue.
	\item If more than $6$ data points are available, take the most recent $6$ and calculate the gradients of the 5 chords in this interval. 
	\item Flag noise if there are 3 or more changes of sign between consecutive gradients.
\end{enumerate}
Once noise is detected, the noisy interval was discarded, the maximal value of $\omega$ was recorded and the process halted. The same was then applied stepping backwards for $\omega < m\omega_{circ}$. This is repeated for all $\lm$ modes with $m\geq 0$ and $\ell + m = \text{even}$, up to some $\ell = \lmax$. The values of the integrals and frequency limits were stored. To minimise duplicate evaluations of the homogeneous solutions $\psi_{\lw}^-$ and $P_{\lw}$ (which are $m$-independent), modes with the same value of $\ell$ but different $m$ were calculated together. 

\textit{Step 2:} Additional $C_{\lmw}^-$ data at intermediate frequencies may be calculated and stored if the desired frequency sampling density is higher than the one used in Step 1.  

\textit{Step 3:} Given the stored $C_{\lmw}^-$ data for all modes up to $\ell=\lmax$, the corresponding $\ell$-modes of the full force $F_{\alpha}^{(\text{full})\ell}$ may be calculated at any point along the orbit using the method of Section~\ref{sec:method_TD_reconstruction}. For a given $\ell$, the terms in the mode sum~\eqref{eq:mode_sum_formula} are calculated by subtracting the regularization terms up to and including $F^{[6]}_{\alpha}$. The self-force may then be approximated by summing the mode-sum up to $\ell = \lmax$. 

Our calculation can be significantly accelerated by making use of parallelization, which we achieved using the OpenMP library \cite{OpenMP_lib}. The calculation of integrals $C_{\lmw}^-$ with different values of $\ell$ and $\omega$ have no common dependencies, which makes these labels ideal for parallelizing over. However, in Step 1 above, the frequencies required are not known in advance, so Step 1 is only parallelized over $\ell$. Step 2 may be parallelized over both $\ell$ and $\omega$ because the frequencies are known in advance, motivating the division between Steps 1 and 2. Parallelization over $m$ is also possible, but this then requires duplicate calculations of the homogeneous solutions, and therefore only advisable if there remain additional unutilized cores after parallelizing over $\ell$ and/or $\omega$. Finally, the calculation of $\lm$-mode contributions to the scalar-field derivatives for Step 3 may also be parallelized over $\ell$, $m$, orbital position or component; we were able to utilize all cores available to us by parallelizing only over orbital position. 

\section{Frequency-domain results}\label{sec:FD_results}

In this section we will examine the effectiveness of the methods we have developed to mitigate the high-frequency noise problem. Satisfied with the results, we will then examine the features of the $C_{\lmw}^-$ spectra.

\subsection{Effect of IBP and tail corrections}\label{sec:IBP_correction_effectiveness}
\begin{figure}
  \centering
  \includegraphics[width=0.49\textwidth]{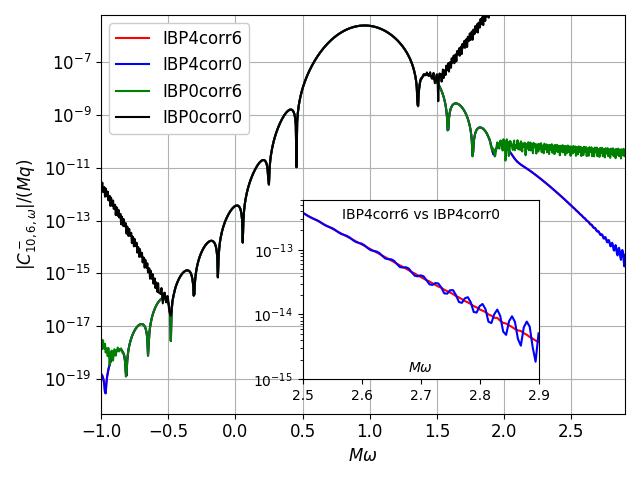}~\\~\\
  \caption[Effect of IBP and corrections]{\label{fig:IBP_correction_comparison} $C_{\lmw}^-$ against frequency for the mode $(\ell, m) = (10, 6)$ and the geodesic with parameters $E = 1.1$ and $\rmin = 4M$, as calculated using the IBP0corr0, IBP0corr6, IBP4corr0 and IBP4corr6 methods. All numerical integrals were truncated at $\rmax = 2000M$. Tail corrections alone are sufficient to delay the onset of noise until further into the tail, but IBP4 is even more effective. Tail corrections provide a small, but non-zero, positive effect when using IBP4.} 
\end{figure}
We begin by investigating the effect of varying the order of IBP and tail correction on the calculation of $C_{\lmw}^-$. Figure~\ref{fig:IBP_correction_comparison} displays $C_{\lmw}^-$ for the mode $(\ell, m ) = (10, 6)$ and the geodesic $E = 1.1$ and $\rmin=4M$, as calculated using different methods. Comparing the IBP0corr0 and IBP0corr6 results confirms that the correction scheme delays the onset of noise and hence allows a greater level of decay to be achieved, in this case gaining approximately 3 orders of magnitude greater decay to the right of the peak. This confirms the utility of the correction scheme, and the successful cancellation between the numerical integral and the tail correction also validates the implementation of the correction. One may confirm that using a lower-order correction produces a smaller, but still positive, improvement. 

If instead we compare IBP0corr6 and IBP4corr0, we see that the IBP4 without tail corrections achieves a greater level of decay than corrections alone, by approximately 3 additional orders of magnitude to the right of the peak in this case. The inset compares IBP4corr0 and IBP4corr6 in the right-hand tail of the spectrum, showing that including the tail corrections introduces further improvement. This improvement, however, is responsible for only a small proportion of the total improvement compared to IBP0corr0, and the effect of including tail corrections is much smaller with IBP4 than with IBP0. Despite this, we believe that continuing to include tail corrections when using IBP is justified on the grounds of improved accuracy (including a reduction in truncation error at intermediate frequencies) and negligible cost. The tail corrections are expressed in terms of gamma functions, which are near-instantaneous to compute compared to the numerical integrals.

\subsection{General features}
\begin{figure*}[t!]%
    \centering  
        \includegraphics[width=0.495\textwidth]{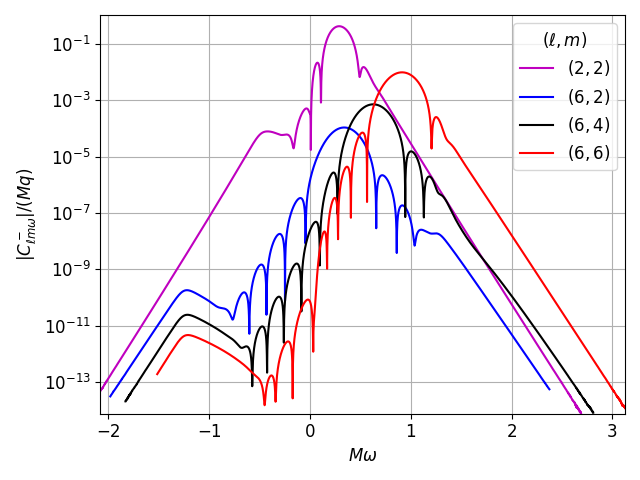}~
    \includegraphics[width=0.495\textwidth]{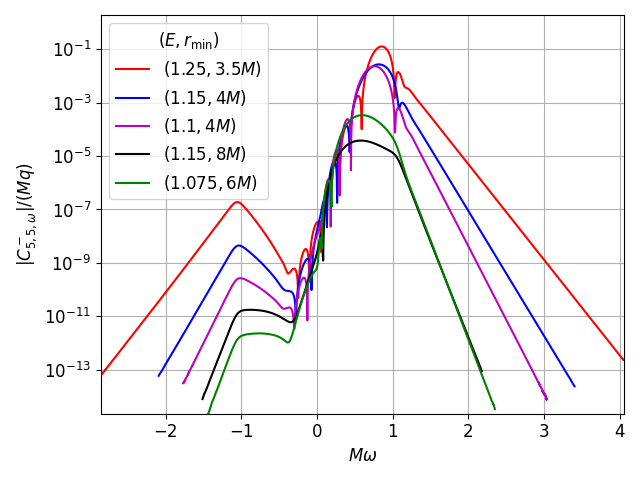}
    \caption{\label{fig:Cminus_examples}Selection of $C_{\lmw}^-$ spectra for a variety of parameters. \textit{Left panel}: different values of $(\ell, m)$ with fixed orbital parameters $E = 1.1$, $\rmin=4M$. \textit{Right panel}: fixed $\ell=m=5$ for different scatter orbits. The IBP4corr6 method was used for frequencies $|M\omega| \geq 0.05$, and IBP04corr6 with $\rsplit = 500M$ was used for frequencies smaller than this. All numerical integrals were truncated at $\rmax = 2000M$. In all cases the displayed frequency range is the maximum one before noise is detected at the endpoints. }
\end{figure*}

The left panel of Fig.~\ref{fig:Cminus_examples} displays sample $C_{\lmw}^-$ spectra for different $\lm$ modes for the geodesic with parameters $E = 1.1$ and $\rmin = 4M$. For fixed $\ell$ and $m\geq 0$, the location of the peak frequency increases in approximate proportion to $m$, while the amplitude at peak also increases with $m$ and decreases with $\ell$. The exponential decay at large $|M\omega|$ is evident.  The right panel of Fig.~\ref{fig:Cminus_examples} instead shows the fixed mode $\ell = m = 5$ for a variety of different orbital parameters. Decreasing the periapsis radius increases the amplitude as expected. Increasing the energy results in a broader spectrum, but only a slight increase in the peak amplitude.

\subsection{Quasinormal modes}

A striking feature in Fig.~\ref{fig:Cminus_examples} is the presence of ``mountains'' in the tail of the spectrum, defined by a sharp triangular peak that interrupts the overall decay trend. As can be seen in the left panel of Fig.~\ref{fig:Cminus_examples}, the location and profile of this feature is roughly independent of $m$ for a given $\ell$. The right panel meanwhile illustrates that this feature occurs in roughly the same location for a wide variety of orbital parameters, although its prominence is variable. These spectral features may occur at either positive or negative frequency, and occasionally both for the same mode.

\begin{figure}
  \centering
  \includegraphics[width=0.495\textwidth]{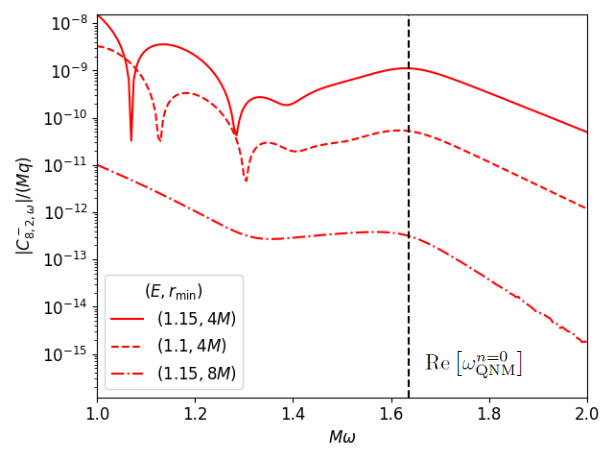}~\\~\\
  \caption[QNM features]{\label{fig:QNM_highlight}$C_{\lmw}^-$ centred on the positive-frequency mountain feature for $(\ell, m) = (8, 2)$ and a selection of orbits. Indicated in vertical dashed line is the real part of the corresponding fundamental quasinormal frequency, ${\rm Re}(\omega_{\rm QNM}^{n=0})\approx 1.63656/M$, which appears to coincide with the peak frequency. The mountain feature becomes more prominent when $\rmin$ is decreased.} 
\end{figure}

These observed behaviors hint at the physical origin of the mountain feature.  Indeed, our investigation reveals that the location of the mountain peak coincides with (plus or minus) the real part of the fundamental quasinormal mode frequency, suggesting these features may be associated with quasinormal excitation of the black hole. We note that for scalar perturbations in Schwarzschild, the quasinormal mode frequencies are independent of $m$, and of the orbital parameters, in line with the above observations. Quasinormal excitation phenomena have previously been observed in self-force calculations for highly eccentric bound orbits \cite{Nasipak:2019hxh, Thornburg:2019ukt}, and subsequently in both gravitational and scalar calculations for scatter orbits \cite{LongBarack2021, BarackLong2022}. 

Figure~\ref{fig:QNM_highlight} illustrates mountains at positive frequency for $(\ell,m)=(8,2)$ and a selection of orbits, also showing the  value of real part of the fundamental quasinormal mode frequency $\omega_{\rm QNM}^{n=0}$. The figure demonstrates the close proximity between the peak and the fundamental quasinormal frequency. It also illustrates the increasing prominence of the quasinormal mode contribution as $\rmin$ decreases and the orbit further penetrates the strong-field region.

\subsection{Zeros in the spectrum}

Another notable feature in Fig.~\ref{fig:Cminus_examples} are locations where $|C_{\lmw}^-|$ drops sharply, becoming very small. These points correspond to frequencies at which both the real and imaginary parts of $C_{\lmw}^-$ appear to vanish simultaneously. Individual zeros of the real and imaginary parts are expected, but it is not a priori clear what physical or mathematical mechanism is responsible for simultaneous zeros. Features suggestive of these zeroes have previously been observed in discrete spectra (see e.g. Fig. 1 in Ref.~\cite{vandeMeent2016}), but we are not aware of any any proposed explanations.

One possible explanation for this phenomenon comes from considering the behaviour of the homogeneous solutions $\psi_{\lw}^{\pm}$ at low frequency. As described in Section~\ref{sec:homsols}, boundary conditions for $\psi_{\lw}^-$ are specified near the horizon and then integrated outwards. For sufficiently small frequencies, there is a region $r \lesssim \omega^{-1}$ where the potential term in Eq.~\eqref{eq:ScalarEOMFD_hom} dominates over the $\omega^2$ term, and $\psi_{\lw}^-$ behaves like a static solution, quickly approaching a (complex-valued) multiple of the polynomially growing real-valued solution $\psi_{\ell,\omega=0}^-$. Likewise, boundary data for $\psi_{\lw}^+$ is specified in the large-$r$ wave-zone, and it generically becomes proportional to the real-valued $\psi_{\ell,\omega=0}^+$ as we move inwards into the potential-dominated region. 

Therefore, if $\omega$ is sufficiently small that a potential-dominated region exists, and if $\rmin \lesssim \omega^{-1}$, then there is a radial range $\rmin < r \lesssim \omega^{-1}$ where the homogeneous solutions may be approximated by (frequency-dependent) complex multiples of the respective static solution. It follows that in this region,  the integrand of Eq.~\eqref{eq:Cminus_def}  can be expressed as a real function of $r$ multiplied by a frequency dependent complex constant. The integrand is largest in the potential-dominated region $\rmin < r \lesssim \omega^{-1}$ on account of the quasi-static polynomial growth of the homogeneous solutions, and thus the integral over this region is expected to provide the dominant contribution to $C_{\lmw}^-$. The real-valued factor of the integrand is oscillatory, and thus the integral of this is an oscillatory function of $\omega$ too; where the real-valued integral vanishes, both real and imaginary parts of $C_{\lmw}^-$ vanish simultaneously.

\section{Self-force results}\label{sec:SF_reconstruction}


We illustrate the calculation of the self-force with the example of the geodesic orbit with parameters $E = 1.1$ and $\rmin=4M$, which is displayed in Fig.~\ref{fig:impact_and_scatter}. The coefficients $C_{\lmw}^-$ were calculated for $\ell$ up to a maximum value $\ell_{\max} = 25$ using the IBP4corr6 method with $\rmax = 2000M$ for frequencies $|M\omega| \geq 0.05$, and IBP04corr6 with $\rsplit = 500M$ and $\rmax = 2000M$ for $|M\omega| < 0.05$. The $\lm$-modes of the scalar field derivatives in the time-domain are obtained by integrating Eqs.~\eqref{eq:FullF_t}-\eqref{eq:FullF_phi} numerically and then constructing $F_{\alpha-}^{\rm(full)\ell}$ using Eq.~\eqref{eq:SSF_sum_over_m}, as described in Sec.~\ref{sec:method_TD_reconstruction}.


As a first test, we validate our code against the analytically known regularization parameters, confirming that the terms in the mode sum \eqref{eq:mode_sum_formula} decay with $\ell$ at the expected rate. We then display the self-force along the orbit. At both stages we compare the results obtained using our frequency-domain (FD) code to those obtained with the time-domain code developed in Refs.~\cite{LongBarack2021, BarackLong2022}, hereafter referred to simply as the time-domain (TD) code.

\subsection{Large-$\ell$ behaviour and code validation}\label{sec:reg_param_tests}

\begin{figure}
  \centering
  \includegraphics[width=0.49\textwidth]{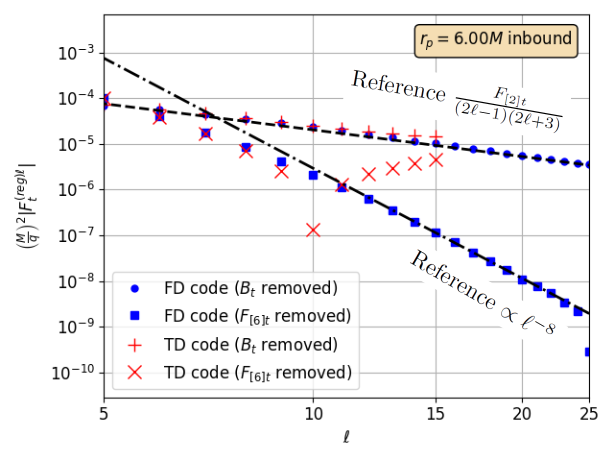}~\\~\\
  \caption[Regplot]{\label{fig:SSF_regplot}Regularized $\ell$-mode contributions to the $t$ component of the self-force at the point $r_p=6M$ along the inbound leg of the orbit with parameters $E = 1.1$ and $\rmin=4M$. The corresponding results from the time-domain code of Refs.~\cite{LongBarack2021, BarackLong2022} are also displayed for comparison for $5 \leq \ell \leq 15$. Two different levels of regularization, subtracting parameters up to and including $B_t$ or $F_{[6]t}$, are displayed. The $B$-regularized and $F_{[6]}$-regularized data agree well at large $\ell$ with the reference lines $\frac{F_{[2]t}}{(2\ell-1)(2\ell+3)}$ (dashed) and $\propto \ell^{-8}$ (dash-dot) respectively, validating our code. Note how the $F_{[6]t}$-regularized TD data becomes noise-dominated already at $\ell\sim 10$, while the corresponding FD data remains faithful down to $\ell\sim 24$. Evidently, the FD calculation is  much more precise.} 
\end{figure}

Figure~\ref{fig:SSF_regplot} displays the regularized $\ell$-mode contributions to the $t$ component of the self-force at the point $r_p=6M$ along the inbound leg of the orbit, with two different levels of regularization applied. In the first set of data, represented by solid circles, the regularization terms involving $A_t$ and $B_t$ have been subtracted from $F_{\alpha}^{\text{(full)}\ell}$, as in the summand of Eq.~\eqref{eq:mode_sum_formula}. At large $\ell$ the terms of this series are known to behave like the higher-order regularization terms in expression~\eqref{eq:higher_order_reg_terms}, decaying as $\ell^{-2}$. Comparison with the reference line representing the first term in that expression confirms that our numerical results have the correct asymptotic behaviour, holding until at least $\ell = 25$.

The convergence of the mode-sum may be accelerated by subtracting the higher-order regularization terms in expression~\eqref{eq:higher_order_reg_terms} from the summand of Eq.~\eqref{eq:mode_sum_formula}. Once all terms up to and including $F_{[6]\alpha}$ have been subtracted, the terms of the mode sum should behave asymptotically as the first neglected term in expression~\eqref{eq:higher_order_reg_terms}, which is expected to decay as $\ell^{-8}$ (with a coefficient we don't currently have). The terms in the mode-sum with parameters up to and including $F_{[6]t}$ removed are represented in Fig.~\ref{fig:SSF_regplot} with solid squares, and good agreement with the reference line proportional to $\ell^{-8}$ is seen all the way until $\ell=25$, when the numerical data begins to deviate from the trend line.

Obtaining the correct asymptotic behaviour is a strong internal check on the accuracy of our code. The $\ell$-mode contributions to the full-force diverge like $\ell$, so to achieve the expected decay rate requires delicate cancellation between the numerically calculated $\ell$-mode and the regularization terms, with a greater degree of cancellation required at larger $\ell$ and when greater numbers of regularization parameters are subtracted. Once $\ell$ becomes large enough, the required degree of cancellation exceeds the precision of the numerical cancellation, and noise is expected to appear. In the example in Fig.~\ref{fig:SSF_regplot}, this is first observed for the $F_{[6]}$-regularized data at around $\ell = 25$.

Additional, external validation is provided by comparison with the TD results, displayed in red in Fig.~\ref{fig:SSF_regplot}. At small $\ell$ there is good agreement, but by $\ell = 10$ the $F_{[6]}$-regularized TD results have visibly broken away from the corresponding FD ones, with the latter continuing to approach the reference lines. By $\ell = 15$, this deviation is visible even in the $B$-regularized modes. Given the superior agreement with the regularization parameters, it is evident that the FD code is significantly more accurate than the TD code at large-$\ell$, at least at this orbital position. This also enables the FD code to reach larger values of $\ell$ before noise appears in the regularized modes, allowing us to use larger values of $\lmax$ than the TD code, and thus reducing the error from truncating the mode-sum at finite $\ell$. 

\subsection{Self-force along the orbit}
\begin{figure*}[t!]%
    \centering
        \includegraphics[width=0.80\textwidth]{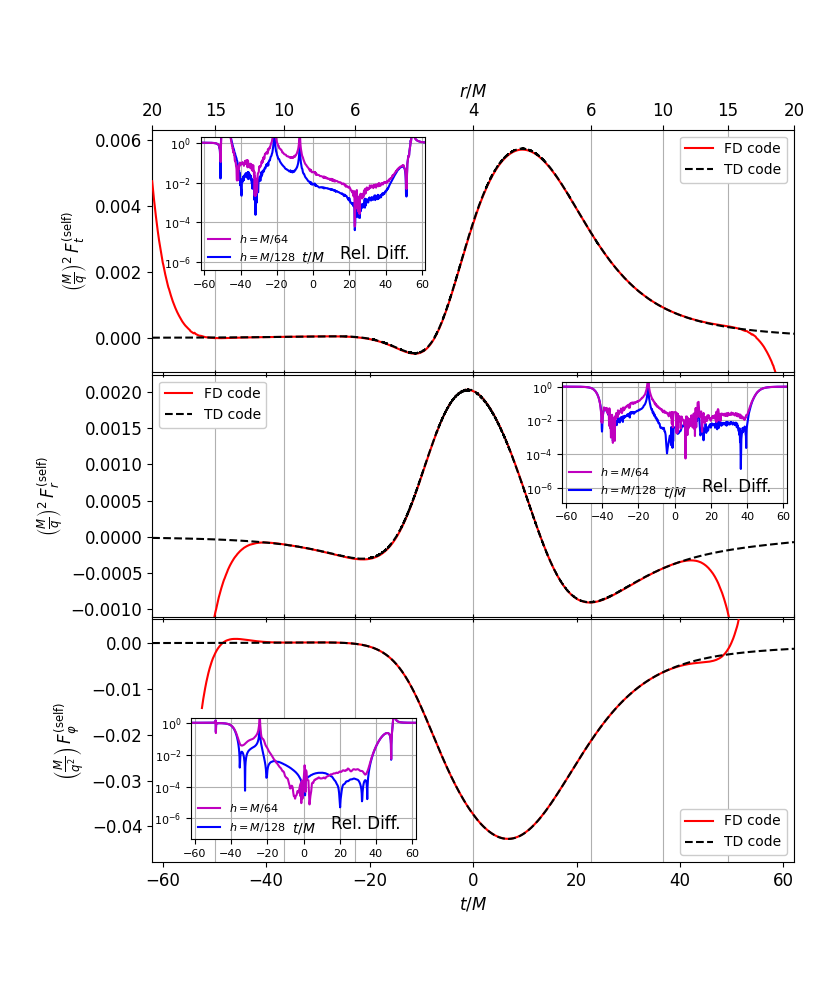}
    \caption{\label{fig:SSF_along_orbit}Components $F_{t}^{\rm(self)}$ (\textit{top}), $F_r^{\rm(self)}$ (\textit{middle}) and $F_{\varphi}^{\rm(self)}$ (\textit{bottom}) of the scalar-field self-force along the orbit with parameters $E = 1.1$ and $\rmin = 4M$, plotted against time $t$ (lower axis) and orbital radius $r$ (upper axis). Results of the frequency-domain code developed in this paper (solid red) are compared to those of the time-domain code developed in Refs.~\cite{LongBarack2021, BarackLong2022} (dashed black, grid spacing $h=M/128$). Periapsis occurs at $t=0$, $r=\rmin=4M$, which is represented by the central vertical line. \textit{Insets:} relative difference between the frequency-domain code and two runs of the time-domain code with different grid spacings $h=M/128$ and $h=M/64$, normalised  by the frequency-domain results. A fiducial choice of $\lmax=15$ and $F_{[6]}$ regularization have been adopted for this test. There is good agreement between the two methods near periapsis, but the accuracy of the frequency-domain code deteriorates rapidly for $r_p \gtrsim 10M$.  }
\end{figure*}

Figure~\ref{fig:SSF_along_orbit} displays the $t$, $r$ and $\varphi$ components of the self-force $F_{\alpha}^{\rm(self)}$ as functions of time $t$ and radius $r$ along our sample orbit. We used $\lmax = 15$, to allow like-to-like comparison with existing TD results, which are also plotted for comparison. The self-force is, predictably, largest in the vicinity of periapsis ($r=\rmin$, $t = 0$), but the peaks are offset from the periapsis position. This behaviour was previously noted for the scalar-field self-force along scatter orbits in Ref.~\cite{BarackLong2022}, and for bound orbits (e.g. Ref.~\cite{Haas07}).

There is a good agreement between the FD and TD codes in the near-periapsis region $4M \leq r \lesssim 10M$ ($|t| \lesssim 37M$), where the results are visually indistinguishable. The insets to Fig.~\ref{fig:SSF_along_orbit} display the relative difference (normalized by the results of the FD code) between the two methods for two different choices of the TD grid spacing $h$. 
With the higher TD resolution ($h=M/128$), the relative difference is between $10^{-3}$ and $10^{-2}$ in the post-periapsis, small radius region for the $t$ and $r$ components, and slightly larger than this shortly before periapsis. For the $\varphi$ component, a tighter agreement of between $10^{-4}$ and $10^{-3}$ is achieved in this region. Near periapsis, the relative difference is sensitive to the resolution used in the TD code, suggesting that the FD code is more accurate here, consistent with our findings in Sec.~\ref{sec:reg_param_tests}. The relative difference gives an estimate of the relative numerical error in the TD calculation, and an upper bound on the numerical error of the FD calculation with this choice of $\lmax$. The closer agreement for $F_{\varphi}^{\rm(self)}$ may be attributed to the greater accuracy of the TD code for this component, which (unlike the $t$ or $r$ components) is obtained from the scalar field itself, without having to take numerical derivatives. 

As we move outwards along the orbit, the agreement between the TD and FD results deteriorates, independently of the TD resolution. For $F_r^{\rm(self)}$ and $F_{\varphi}^{\rm(self)}$ this occurs at approximately $r = 10M$ ($t\sim 37M$), but agreement is maintained until around $r = 15M$ ($t\sim 50M$) for $F_t^{\rm(self)}$. Referring back to the main plots in Fig.~\ref{fig:SSF_along_orbit}, it is clear that the FD  method is responsible, breaking sharply away from the smoothly decaying curve obtained using the TD method. An examination of the $\ell$-mode contributions to the self-force at these larger radii shows that the regularized $\ell$-modes in the tail of the mode sum cease to decay, and instead begin to blow up rapidly with $\ell$. As the radius is increased, the problem affects successively lower values of $\ell$. This phenomenon is associated with error messages from the numerical integrator, indicating that the inverse Fourier integrals \eqref{eq:FullF_t}-\eqref{eq:FullF_phi} cannot be evaluated to the requested relative error tolerance of $10^{-8}$.

The above results are reassuring, if mixed. The successful regularization tests and agreement with TD results validate our new code, and eliminate the possibility of any serious errors in the calculation of FD quantities such as the $C_{\lmw}^-$. We have illustrated the higher accuracy of the new FD code at small radii (compared to the existing TD implementation), where it exhibits superior large-$\ell$ performance and greater accuracy in the summed force. However, as $r$ is increased, the code quickly loses accuracy at large $\ell$. 
In practice, this means that, without directly addressing the problem, we are limited to $\lmax$ values that must be made smaller with increasing $r$, at a cost of increased truncation error. 

In Sec.\ \ref{sec:cancellation_problem} we discuss the cause of the problem at large $r$ and possible remedies for it. Since we are yet to develop a complete satisfactory cure, we proceed here with a temporary solution, involving a procedure for an adaptive adjustment of $\lmax$ as a function of $r$ along the orbit.

\subsection{Adaptive truncation of $\ell$-mode summation}\label{sec:SSF_dyn_ltrunc}
\begin{figure*}%
    \centering
        \includegraphics[width=0.80\textwidth]{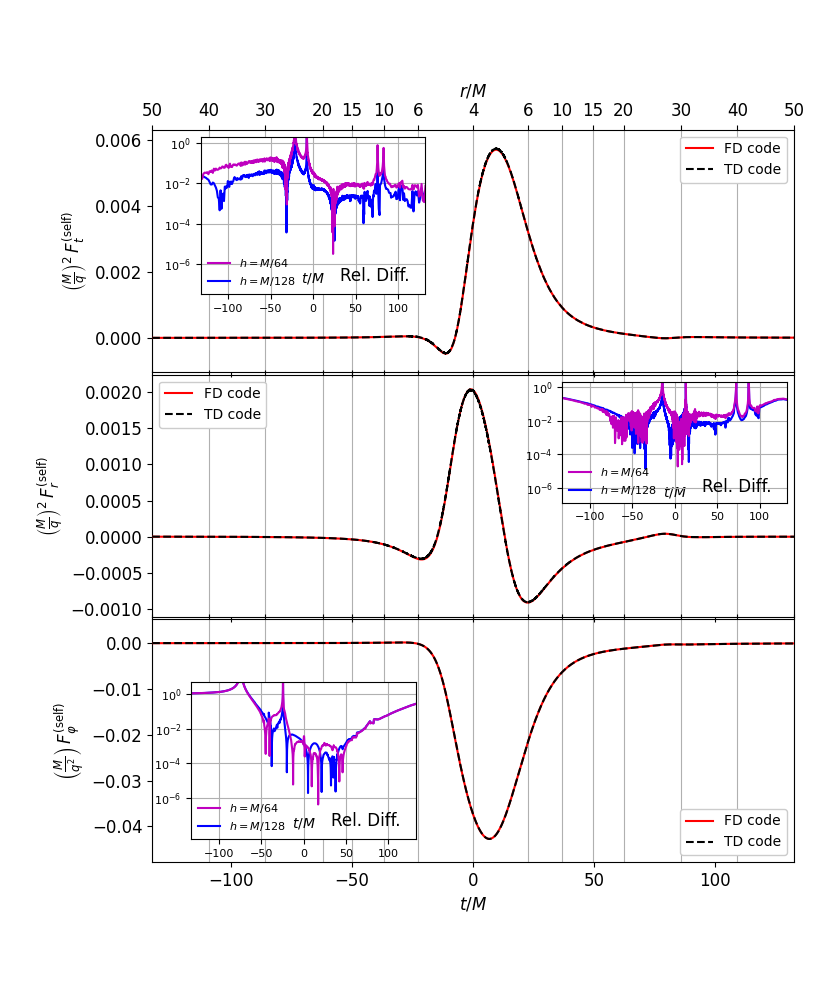}
    \caption{\label{fig:SSF_dyn_ltrunc}
    Same as Fig.\ \ref{fig:SSF_along_orbit}, after applying the adaptive mode-sum truncation procedure described in the text. The time-domain data continues to use $\lmax = 15$. The improvement at large $r$ is manifest.
}
\end{figure*}
To detect when the terms of the mode-sum begin to rapidly lose accuracy, the following algorithm was applied. First, regularized $\ell$-modes $F_{\alpha}^{\rm(reg)\ell}$ are calculated and added for $0 \leq \ell \leq \ell_{\rm min}$, where $\ell_{\rm min}$ takes some pre-selected value. Additional modes are then added so long as at least one of the following two criteria are met: Either $\left|F_{\alpha}^{\rm(reg)\ell}\right| < \sigma_1\left|F_{\alpha}^{\rm(reg)\ell-1}\right| $; or there is precisely one change of sign among the successive $\ell$-modes $F_{\alpha}^{\rm(reg)\ell-3}$, $F_{\alpha}^{\rm(reg)\ell-2}$, $F_{\alpha}^{\rm(reg)\ell-1}$ and  $F_{\alpha}^{\rm(reg)\ell}$,  and, in addition, $\left|F_{\alpha}^{\rm(reg)\ell}\right| < \sigma_2 \>\text{max}\left(\left|F_{\alpha}^{\rm(reg)\ell-2}\right|, \left|F_{\alpha}^{\rm(reg)\ell-3}\right| \right)$.
The second condition is designed to allow change-of-sign features to pass through the filter without triggering the truncation mechanism. Repeated changes of sign are taken to indicate noise or other difficulties in the $\ell$-modes, and treated as a trigger for truncation. The constants $\sigma_{1,2}$ are safety factors to prevent the truncation mechanism from being falsely triggered by any legitimate small-scale features, such as ``bounce-back" after a change of sign. Values of $\sigma_1 = 1.1$ and $\sigma_2 = 2$ were adopted. A minimum number of $\ell$-modes, $\ell_{\rm min}$, are always included, so that the truncation mechanism is not triggered by transient small-$\ell$ behaviour. A piecewise value $\ell_{\rm min} = 10$ for $r_p \leq 10M$ and $\ell_{\rm min}= 5$ for $r_p > 10M$ was selected. The maximum possible value of  $\lmax$ is set by the largest value of $\ell$ for which $C_{\lmw}^-$ data is available, which in this paper was $\lmax=25$. 

$F_{[6]}$-regularized $\ell$-mode contributions to the self-force were calculated in advance up to $\ell = 25,\>20$ and $15$ for $r_p \leq 6M$, $6M < r_p \leq 15M$ and $r_p > 15M$ respectively, at points along the orbit which are uniformly spaced in $\chi$ between $r_p = 50M$ (inbound) and $r_p = 50M$ (outbound). The truncation algorithm was then applied to determine the appropriate value of $\lmax$ at each location. The resulting self-force along the orbit is displayed in Fig.~\ref{fig:SSF_dyn_ltrunc}, with the TD results (still using $\lmax =  15$) for comparison.

We observe that the use of adaptive truncation of the mode sum prevents any visible blow-up in the FD calculation, out to at least $r = 50M$. This represents a significant improvement compared to the fiducial $\lmax = 15$ test in Fig.~\ref{fig:SSF_along_orbit}. The insets to Fig.~\ref{fig:SSF_dyn_ltrunc} once again display the relative difference between the FD results and two TD runs with different resolutions ($h = M/128$ and $h = M/64$). For the $t$ component of the force, the relative difference is sub-1\% at late times, and $O(1)$-$O(10)\%$ at early times, with significant sensitivity to the TD resolution used. This once again suggests that the FD code is more accurate here, even at $r = 50M$. For the $r$ and $\varphi$ components, however, the relative difference is broadly insensitive to the TD resolution used, and the errors increase with radius, becoming $O(10)\%$ at late time, and even larger at very early time. To pin down the dominant source of numerical error in these domains would require a more detailed analysis of both FD and TD codes, which we have not performed here. We suspect the large-$\ell$ truncation error in the FD code is dominant at large $r_p$.

\section{Scattering angle }\label{sec:scatter_angle_calc}
\subsection{Method}\label{sec:scatter_angle_method}
We will now illustrate a full calculation of the self-force correction to the scatter angle, $\delta\varphi^{(1)}$ at fixed $(b, v)$. We do this for our standard background geodesic with parameters $E = 1.1$ and $\rmin = 4M$ (exactly), corresponding to 
\begin{align}
    b \approx 10.401M, \quad v \approx 0.41660.
\end{align}
In Eq.~\eqref{eq:1SF_scatter_angle} we gave $\delta\varphi^{(1)}$ in terms of an integral of self-force quantities along the orbit. It is customary to split $\delta\varphi^{(1)}$ into conservative and dissipative contributions, which is conveniently done by making use of the symmetries
\begin{align}
    F_{\alpha}^{\perp\rm(cons)}(\chi) = -F_{\alpha}^{\perp\rm(cons)}(-\chi),\\
    F_{\alpha}^{\perp\rm(diss)}(\chi) = F_{\alpha}^{\perp\rm(diss)}(-\chi),
\end{align}
for $\alpha = t, \varphi$. One finds 
\begin{align}
    \delta\vf^{(1)}_{\rm(cons)} = \int_{0}^{\chiinf} \Big[\mathcal{G}^{\rm(cons)}_E&(\chi)F_t^{\bot\rm(cons)}(\chi) \label{eq:1SF_scatter_angle_cons} \\ \nonumber &- \mathcal{G}^{\rm(cons)}_L(\chi)F_{\varphi}^{\bot\rm(cons)}(\chi)\Big]\tau_{\chi}d\chi,
\end{align}
\begin{align}
    \delta\vf^{(1)}_{\rm(diss)} = \int_{0}^{\chiinf} \Big[\beta_E F_t^{\perp\rm(diss)} - \beta_L F_{\vf}^{\perp\rm(diss)} \Big]\tau_{\chi} d\chi,\label{eq:1SF_scatter_angle_diss}
\end{align}
where the functions $\mathcal{G}^{\rm(cons)}_{E/L}(\chi)$ and constants $\beta_{E/L}$ are given explicitly in Eqs.~(85) and (117) of Ref.~\cite{BarackLong2022}, respectively. The total scatter angle correction is then 
\begin{align}
\delta\vf^{(1)} = \delta\vf^{(1)}_{\rm(cons)} + \delta\vf^{(1)}_{\rm(diss)}.
\end{align}

We implement Eqs.~\eqref{eq:1SF_scatter_angle_cons} and \eqref{eq:1SF_scatter_angle_diss} numerically in a Python script, which takes as input the total, unprojected, self-force $F_{\alpha}^{\rm(self)}$ at $8N$ discrete locations, $\chi_n := n h_{\chi}$ for $n = \pm 1, \pm 2,..., \pm 4N$. Here $N$ is an integer, $h_{\chi}$ is the spacing of the $\chi$ sample, and $\chi_{\rm max} := 4N h_{\chi}$ is the value of $\chi$ where the integrals will be truncated. (The point $\chi = 0$ is not initially included in our sample because, although the entire integrand is bounded at $\chi = 0$, the factors $\mathcal{G}^{\rm(cons)}_{E/L}$ are individually singular there; instead, we obtain the value at $\chi=0$ via extrapolation.) Given the unprojected self-force data, the script calculates the projection of the self-force orthogonal to the 4-velocity using Eq.~\eqref{eq:SSF_perp_def}, and then it separates it into conservative and dissipative pieces $F_{\alpha}^{\perp\rm(cons)}$ and $F_{\alpha}^{\perp\rm(diss)}$ using Eqs.~\eqref{eq:cons_def}-\eqref{eq:adv_from_ret}. 

The integrands in Eqs.~\eqref{eq:1SF_scatter_angle_cons} and \eqref{eq:1SF_scatter_angle_diss} are then constructed at the $4N$ positions in $0 < \chi \leq \chi_{\rm max}$, and extended to $\chi = 0$ via extrapolation. The scatter-angle integrals (truncated at $\chi = \chi_{\rm max}$) are evaluated using Simpson's $1/3$ rule, which requires an odd number of data points. Because $4N+1$ data points are available, 2 estimates of the integral may be obtained, one using step width $h_{\chi}$, and another using step width $2h_{\chi}$. The former is used as the best estimate for the scatter angle, while the difference
\begin{align}
    \epsilon = \frac{1}{15}\left(\delta\vf^{(1)}(2h_{\chi}) - \delta\vf^{(1)}(h_{\chi})\right)
\end{align}
provides an estimate of the quadrature error.

\subsection{Results}\label{sec:scatter_angle_results}
The self-force was calculated out to $r_p = 50M$ along both inbound and outbound legs of the orbit, with $\chi$-spacing $h_{\chi} = \chi_{50}/1024$, where $\chi_{50}\approx 2.0776$ is defined by $r_p(\chi_{50}) = 50M$. This choice of $h_{\chi}$ is found to produce a quadrature error much smaller than other sources of error in our calculation (see below). The integrals \eqref{eq:1SF_scatter_angle_cons}~and ~\eqref{eq:1SF_scatter_angle_diss} were truncated at $\chi_{\rm max} = \chi_{50}$. The resulting estimates for the scatter angle correction, expressed to 5 significant figures, were
\begin{align}
    \delta\vf^{(1)}_{\rm (cons)} = -1.5032, \quad\quad
    \delta\vf^{(1)}_{\rm (diss)} =2.7034 \nonumber \\
   \text{(FD, $r_{\max}=50M$)},
    \label{eq:scatter_angles}
\end{align}
with respective estimated quadrature errors
    $\epsilon_{\rm(cons)} \approx -1.1\times10^{-7}$ and  $\epsilon_{\rm(diss)} \approx -7.9\times10^{-7}$.

For comparison, we apply the same method (with the same values of the $\chi$-spacing $h_{\chi}$ and truncation point $\chi_{\rm max}$) to the self-force data obtained using the TD code of Refs.~\cite{LongBarack2021, BarackLong2022}. This approach yields estimates of 
\begin{align}
    \delta\vf^{(1)}_{\rm(cons)} = -1.5309, \quad\quad
    \delta\vf^{(1)}_{\rm(diss)} = 2.6950 \nonumber \\
       \text{(TD, $r_{\max}=50M$)},\label{eq:scatter_angles_mixed}
\end{align}
The difference relative to our estimates in Eq.~\eqref{eq:scatter_angles} is approximately $1.8\%$ ($0.31\%$) in the conservative (dissipative) pieces. This discrepancy is significantly larger than the quadrature errors estimated above. In Sec.~\ref{sec:SF_reconstruction} it was found that the disagreement between the FD and TD self-force values was greatest at large radius, and thus it is expected that the large-radius portions of the integrals \eqref{eq:1SF_scatter_angle_cons} and \eqref{eq:1SF_scatter_angle_diss} contribute most significantly to the discrepancy in the scatter angle estimates. This may be investigated by truncating the integrals \eqref{eq:1SF_scatter_angle_cons} and \eqref{eq:1SF_scatter_angle_diss} at a smaller value of $\chi$ and examining the effect on the discrepancy. Using the same value of $h_{\chi}$ but a reduced truncation position $\chi_{\rm max} = 976h_{\chi}$ (corresponding to $r_p \approx 29.8M$), the FD self-force data provided estimates $-1.4515$ ($2.6382$) for the conservative (dissipative) piece of the scatter angle correction, compared to $-1.4627$ ($2.6298$) using the TD self-force data. The relative difference between the conservative parts has reduced significantly to $0.77\%$, but remains broadly unchanged at $0.32\%$ for the dissipative part. This supports the idea that, in our tests, the discrepancy between the FD and TD conservative scatter angles is dominated by the loss of accuracy in the FD self-force at large radius, but the (smaller) discrepancy in the dissipative pieces may have a different root cause. 

This test highlights another significant source of error in the estimates \eqref{eq:scatter_angles}, that which arises from truncating the integrals \eqref{eq:1SF_scatter_angle_cons} and \eqref{eq:1SF_scatter_angle_diss} at $\chi < \chi_{\infty}$. To quantify this error, we change integration variable in Eq.~\eqref{eq:1SF_scatter_angle_cons},
\begin{align}
    \delta\vf^{(1)}_{\rm(cons)} = \int_{\rmin}^{\infty} \label{eq:1SF_scatter_angle_cons_rp}\Big[\mathcal{G}^{\rm(cons)}_E&(r_p)F_t^{\bot\rm(cons)}(r_p) \\ \nonumber &- \mathcal{G}^{\rm(cons)}_L(r_p)F_{\varphi}^{\bot\rm(cons)}(r_p)\Big]\frac{dr_p}{\dot{r}_p},
\end{align}
and note that the self-force components decay like $r_p^{-3}$ as $r_p\rightarrow\infty$ \cite{BarackLong2022}. 
A simple asymptotic analysis shows that 
the full integrand in Eq.~\eqref{eq:1SF_scatter_angle_cons_rp} decays like $r_p^{-3}$. Hence, when Eq.~\eqref{eq:1SF_scatter_angle_cons} is truncated at $\chi_{\rm max} < \chiinf$, the resulting truncation error behaves as $r_{\rm max}^{-2}$, where $r_{\rm max} = r_p(\chi_{\rm max})$. The same decay rate applies for the dissipative piece also. 

Based on this reasoning, the relative truncation error in the scatter angle corrections may be approximated by $(\rmin/r_{\rm max})^2$. For truncation at $r_{\rm max} = 50M$, this would suggest an error of approximately $(4/50)^2 = 6.4\times10^{-3}$. This is, of course, a very crude estimate, because the integrand is not a strict power-law function, and may also exhibit changes of sign. To better estimate the true truncation error in our calculation, we may compare to the values obtained in \cite{BarackLong2022}, where use was made of a larger truncation radius, with an analytic approximation for the large-$r$ tail of the integral. This more precise calculation gave 
\begin{align}
    \delta\vf^{(1)}_{\rm(cons)} = -1.5957 \pm 0.0023,\nonumber \\
    \delta\vf^{(1)}_{\rm(diss)} = 2.7612 \pm 0.0026 \nonumber\\
        \text{(TD, $r_{\max}=\infty$)},
    \label{eq:scatter_angle_olly}
\end{align}
suggesting relative errors of approximately $4.1\%$ and $2.5\%$ in the truncated TD estimates \eqref{eq:scatter_angles_mixed}. 
This is significantly larger than our crude $\sim 0.6\%$ estimate. 

In summary, we see that the dominant source of error in our FD value (\ref{eq:scatter_angles}) is the large-$r$ truncation of the orbital integration. 
Mitigating this truncation error is challenging within our current FD framework. As we have discussed, increasing the truncation radius results in significant loss of accuracy. Indeed, even truncating at $r_p = 50M$ we found that the error associated with the large-$r$ self-force may be as much as half the size of the truncation error for the conservative piece. Attempting to extrapolate the self-force to large radii is likewise more challenging for our FD code than it was for the TD code of Refs.~\cite{LongBarack2021, BarackLong2022}. If a tail is fitted to inaccurate large-radius self-force data, the accuracy of the extrapolation will be fundamentally limited. 
It is for this reason we find fitting a tail to our self-force data impractical. It is clear that high-precision scatter angle calculations will require improvements at large radius. 

The next section will explore more deeply the reasons for our problems at large $r$, and suggest mitigations.  

\section{Cancellation problem}\label{sec:cancellation_problem}
\begin{figure}[h!]
  \centering
  \includegraphics[width=0.49\textwidth]{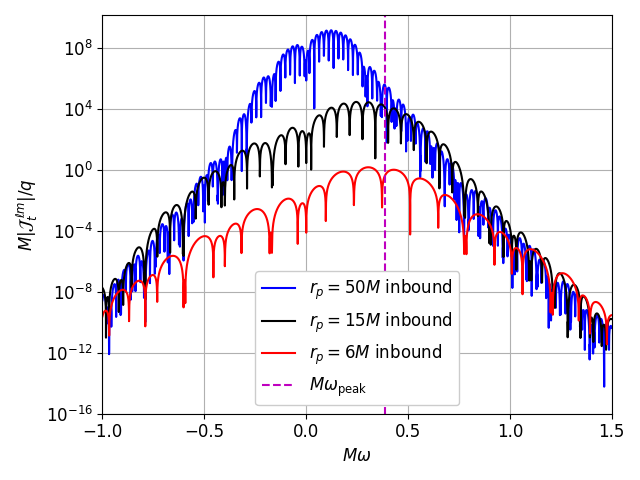}~\\~\\
  \caption[Fourier integrand plot]{\label{fig:Fourier_integrand_plot}Integrand $\mathcal{J}^{\lm}_t$ [as defined in Eq.~\eqref{eq:Fourier_integrand_J}] plotted against $\omega$ for the mode $(\ell, m) = (10, 2)$ at selected radii along the inbound leg of the orbit with parameters $E = 1.1$ and $\rmin=4M$. Also plotted is a vertical line representing the frequency $\omega_{\rm peak}$ at which $C_{10,2,\omega}^-$ peaks. The peak value of the integrand grows rapidly and is increasingly shifted towards $\omega = 0$ as the radius is increased, while the frequency of the oscillations also increases.} 
\end{figure}
We seek to understand the origin of the observed loss of accuracy at large $r$ and large $\ell$. 
To this end, consider the integrand of the Fourier integral $\Phi^{\lm-}_t$ in Eq.~\eqref{eq:FullF_t} as an example. In order to reconstruct $F_{t-}^{\rm(full)\ell}$ using Eq.~\eqref{eq:SSF_sum_over_m}, the real-valued integrand of interest is
\begin{align}
    \mathcal{J}^{\lm}_t(\omega) := -\text{Re}\SB{\frac{i\omega}{r_p}C_{\lmw}^-\psi_{\lw}^- e^{-iwt_p}Y_{\lm}\left(\frac{\pi}{2},\vf_p\right)}\label{eq:Fourier_integrand_J}.
\end{align} In Fig.~\ref{fig:Fourier_integrand_plot} we plot this quantity as a function of frequency at selected radii along the inbound leg of the orbit for the mode $(\ell, m) = (10, 2)$. The peak of $\mathcal{J}^{\lm}_t$ is seen to shift away from that of $C_{\lmw}^-$ and towards $\omega = 0$, where the homogeneous solution factor, $\psi_{\lw}^-(r_p)$, peaks. This offset is small at small radii, but the peak grows rapidly and becomes increasingly shifted towards $\omega = 0$ as $r_p$ is increased, due to the quasi-static growth of the homogeneous solution, $\psi^-_{\ell,\omega=0} \sim r^{\ell+1}$, at relatively small frequencies $\omega \ll \ell r^{-1}$. At the same time, the integrand oscillates at an increasing rate due to the factors $\psi_{\lw}^-(r_p)$ and $e^{-i\omega t_p}$, which have phases $\sim \omega r_*^p$ and $\omega t_p$ respectively. 
The same behavior is observed in the $r$ and $\vf$ components too. 

The $\ell$-modes of the self-force, however, \textit{decay} with radius. The conclusion is that there must be an increasing degree of cancellation in the Fourier integrals as the radius is increased. This increasing cancellation in the time-domain reconstruction due to the quasi-static growth of the EHS was first noted in Ref.~\cite{vandeMeent2016}, which studied the gravitational self-force on particles moving along eccentric, bound, Kerr geodesics. In that paper the author explained that the EHS modes with relatively small frequency exhibit unphysical amplitude variations along the orbit, by an amount that grows exponentially in $\ell$. In the case of our scatter orbits, the unphysical growth of the EHS is encapsulated by Eq.~\eqref{eq:EHS_growth}, which makes the exponential dependence on $\ell$ clear.

\begin{figure}
  \centering
  \includegraphics[width=0.49\textwidth]{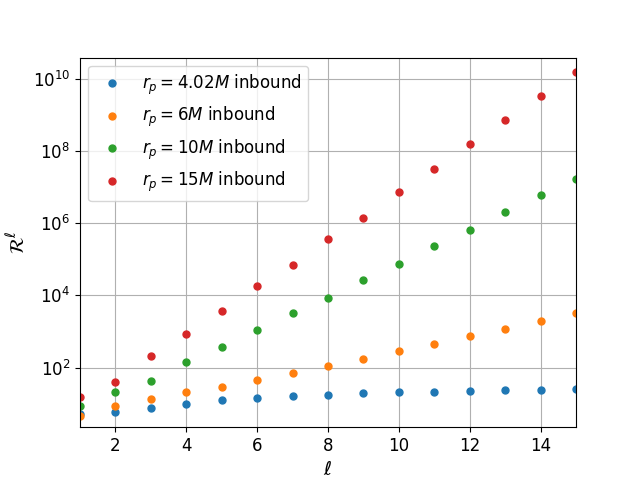}~\\~\\
  \caption[Degree of cancellation]{\label{fig:degree_of_cancellation}Degree of cancellation $\mathcal{R}^{\ell}$ [as defined in Eq.~\eqref{eq:DOC_R}] in the calculation of the Fourier integral $\Phi^{\lm-}_{t}$ at selected points along the inbound leg of the orbit with parameters $E = 1.1$, $\rmin=4M$. The degree of cancellation (and hence loss of precision) increases with $r_p$, and exponentially in $\ell$.  } 
\end{figure}

The cancellation in the Fourier integral amplifies the error in the numerically calculated integrand, and we refer to this phenomenon as the \textit{cancellation problem}. Once the loss of precision exceeds the precision of the underlying frequency-domain calculations, the cancellation in the numerical Fourier integrals cannot occur and our calculated self-force loses accuracy and begins to blow-up. The degree of cancellation (and hence the resulting loss of precision) may be quantified for the case of the $t$ component by
\begin{align}
    \mathcal{R}^{\ell} := \max_{m} \left(\frac{\|\mathcal{J}^{\lm}_t\|_{1}}{\left|\text{Re}\left(\Phi^{\lm-}_tY_{\lm}\right)\right|}\right),\label{eq:DOC_R}
\end{align}
where 
\begin{align}
    \|f\|_{1} := \displaystyle\int_{-\infty}^{+\infty} |f(\omega, r_p)|\>d\omega
\end{align}
is the $L^1$-norm over frequency at fixed orbital position. This quantity is plotted in Fig.~\ref{fig:degree_of_cancellation}, demonstrating the increased cancellation with $r_p$, and in particular reproducing the expected exponential growth with $\ell$.

In Refs.~\cite{vandeMeent2016} and \cite{vandeMeent2018} the author managed the cancellation problem by using arbitrary precision arithmetic, calculating frequency-modes at sufficiently high precision to ensure the desired level of accuracy even after the loss of precision during time-domain reconstruction. The key downside of this approach is the significantly increased computational cost that comes with higher precision arithmetic, which is particularly undesirable for hyperbolic orbits given the already reduced efficiency of frequency-domain methods in this regime. There are several other reasons why this approach is less effective in the scatter problem. First, we have seen that the scatter problem introduces a new source of error, arising from the truncation of the normalization integrals $C_{\lmw}^-$ at a finite radius. To benefit from improved-precision arithmetic would require a commensurate reduction in this truncation error, which, in turn, is likely to demand the use of new techniques, such as the derivation of higher-order IBP rules or the use of compactification \cite{Macedo2022, MacedoLongBarackinprog}, with their own challenges and costs to bear. Second, achieving very high accuracy in the interpolation of $C_{\lmw}^-$ will inevitably require the calculation of a larger number of frequency modes to use as data nodes, introducing extra cost and compounding the first issue. This is in contrast to the discrete-frequency bound case, where approximately the same number of frequencies are required even at radii where the cancellation problem is more severe.

We see several possible directions for mitigating the cancellation problem in future work. The most straightforward approach is to refine our existing double-precision code to improve the precision of the various frequency-domain quantities that go into the Fourier integrals, and hence delay the onset of breakdown. For example, adaptive placement of interpolation nodes for $C_{\lmw}^-$ may be used to reduce interpolation error at those frequencies that contribute strongly to the cancellation, while reducing sample density to conserve computation time at frequencies in the tail of the spectrum, where the error requirement is less stringent. Another option is the use of semi-analytical results, based around small-$\omega$ expansions of the homogeneous solutions, to improve the accuracy of $C_{\lmw}^-$ around $\omega = 0$. A more ambitious objective would be to use small-$\omega$ expansions of the homogeneous solutions to better understand the nature of the cancellation, with an aim to achieve part of the cancellation analytically.

In parallel to these direct mitigations of the cancellation problem, one should develop analytical approximations for the large-$r$ tail of the self-force, which would reduce the need for a numerical calculation at large radius. This would also be beneficial for the TD approach, which also suffers from a loss of accuracy and efficiency at large radii. We are currently working to obtain such auxiliary analytical expressions.  

It should be reminded here that the cancellation problem is an inherent feature of the EHS approach, where unphysical, growing homogeneous solutions are extended into the source region. The problem would not occur in a calculation based on the standard variation-of-parameters formula, if a way was found to enable an efficient reconstruction of the TD field at the particle despite the Gibbs phenomena. Ultimately, therefore, a full satisfactory solution to the cancellation problem might need to involve a departure from the usual EHS approach. Alternatives would have to tackle the Gibbs phenomenon head-on. Ideas to be explored involve the application of spectral filtering to improve the convergence of the Fourier integral near the particle \cite{GibbsResolution}; and/or the use of  extrapolation to estimate the one-sided limits of the field derivatives based on values calculated away from the particle location, where the impact of the Gibbs phenomenon is less severe.   

\section{Discussion}\label{sec:discussion}
In this work we developed and tested a frequency-domain approach to self-force calculations for scattering orbits. As discussed in Sec.~\ref{sec:EHS_section}, the standard EHS approach cannot be applied in the external region $r \geq r_p(t)$ when dealing with scatter orbits, leading us to use a one-sided mode-sum regularization approach, which required only one-sided scalar-field derivatives at the particle's location, taken from the internal region $r \leq r_p(t)$. The time-domain scalar field may still be reconstructed from an EHS in this internal region but, as explored in Sec.~\ref{sec:truncation_problem}, the numerical evaluation of the normalization integrals $C_{\lmw}^-$ is complicated by the need to truncate a slowly convergent integral over the non-compact radial extent of the orbit, a novel challenge unique to the scatter problem. To mitigate this, the \textit{tail correction} and \textit{integration by parts} techniques were developed in Sec.~\ref{sec:radial_integral} to minimise the truncation error that occurs, and the success of these approaches is illustrated in Sec.~\ref{sec:IBP_correction_effectiveness}. The development of these techniques, particularly integration by parts, forms one of the central results of this work. It was critical for enabling the self-force calculation that followed.

In Sec.~\ref{sec:SF_reconstruction} we validated our self-force results using comparisons with analytically known regularization parameters at large-$\ell$, and with the TD code of Refs.~\cite{LongBarack2021, BarackLong2022}. Based on the superior agreement with the regularization parameters at large-$\ell$, and comparisons to different resolution runs of the TD code, we concluded that our FD code is capable of superior accuracy in the near-periapsis region of the orbit. However, it was also found that the FD code suffers a severe loss of accuracy as the radius is increased along the orbit. Although partially tamed by the use of adaptive ($r$-dependent) truncation of the mode-sum, the accuracy of the FD code still decreases relative to the TD code as radius is increased, and in Sec.~\ref{sec:scatter_angle_calc} it was observed that the poor large-$r$ behaviour of the FD code was the limiting factor in the accuracy of our scatter angle calculation. 

In Sec.~\ref{sec:cancellation_problem} the loss of accuracy at large $r$ was explained in terms of a previously observed  cancellation problem that occurs when reconstructing the time-domain field from frequency modes of an EHS for highly eccentric orbits \cite{vandeMeent2016, vandeMeent2018}. Several possible solutions and mitigations were discussed, including ways to reduce the severity of the problem within the current EHS framework, circumventing the problem altogether with non-EHS FD techniques, and supplementing the numerical calculation with analytical expansions of the self-force at large radius. 

 The FD and TD codes have thus far been running on different computing platforms, with different numbers of cores and single-thread performances, preventing direct runtime comparisons. Furthermore, there is further room to optimise the FD code, and so comparisons at this time may not accurately capture the code's potential. For example, the equal-$\chi$ sampling of the self-force used for the scatter angle calculation in Sec.~\ref{sec:scatter_angle_calc} is not optimal; the FD code is capable of calculating the self-force at any requested orbital location, and thus may be used as input to an adaptive integration scheme, which minimises the number of self-force evaluations for a given quadrature error in the scatter angle integrals. The FD algorithm is much less costly in terms of RAM usage and data storage. We are planning to conduct a more direct assessment of computational saving in future work.

The ultimate aim of our program is to calculate gravitational self-force corrections to the scatter angle across the geodesic parameter space. 
The FD method developed in this work represents a step towards this goal, but the method remains limited by the cancellation problem. To resolve this issue, we intend to investigate two parallel strands. First, we will develop large-$r$ analytical expressions for the self-force, with the aim of replacing the numerical self-force data at large radius. This would be also benefit the parallel TD effort.  Second, we will consider alternative approaches to the use of EHS in the frequency-domain. This would remove the cancellation problem altogether, and potentially enable the use of two-sided regularization, typically required for radiation-gauge approaches to the gravitational self-force problem \cite{Pound:2013faa}. Alternatives to EHS would also benefit self-force calculations along highly-eccentric bound orbits, which also suffer from the cancellation problem.

In the longer term, we intend to explore the applicability and performance of our FD method for weak-field (large $r_{\rm min}$) orbits, where comparison with analytical post-Minkowskian calculations provide interesting opportunities already in the scalar-field problem \cite{Barack:2023oqp}. Once a solid FD method is at hand, we will turn to tackle the gravitational problem via the radiation-gauge formalism of Ref. \cite{Pound:2013faa}, which would build on the scalar-field framework in a natural way. 

\begin{acknowledgments}
We are deeply grateful to Oliver Long for producing and providing comparison data from his time-domain code, and for his useful comments and advice throughout this work. We also thank Maarten van de Meent and Niels Warburton for useful discussions.
CW acknowledges support from EPSRC through Grant No. EP/V520056/1.
This work makes use of the Black Hole Perturbation Toolkit.
\end{acknowledgments}

\appendix
\section{Expansion of geodesic quantities at large radius}\label{app:geodesic_expansion_coeffs}
We collect here expressions for the coefficients that occur in the large-radius expansions of the geodesic quantities in Eqs.~\eqref{eq:texp} and \eqref{eq:phiexp}. To obtain large-$r$ expansions for $t_p(r)$ and $\varphi_p(r)$ along the outbound leg of the orbit, we rewrite the equations of motion in terms of the parameter $r$, and then expand in powers of $1/r$. In order to do this, we first expand 
\begin{align}
	\SP{\frac{dr_p}{d\tau}}^{-1} =\sum_{n=0}^{\infty}U_n\left(\frac{2M}{r}\right)^n,\label{eq:urinv_exp_app}
\end{align}
where the first few coefficients work out to be
\begin{align}
	U_0&= \frac{1}{\sqrt{E^2-1}},\\
	U_1 &= -\frac{1}{2 \left(E^2-1\right)^{3/2}},\\
	U_2 &= \frac{3 - 4 \tilde{L}^2 + 4 E^2 \tilde{L}^2}{8 (E^2 - 1)^{5/2}}, \\
	U_3 &= \frac{-5 + 4\tilde{L}^2 + 4 E^2 \tilde{L}^2 - 8 E^4 \tilde{L}^2}{16 (E^2 - 1)^{7/2}},\\
	U_4 &=\frac{1}{128 \left(E^2-1\right)^{9/2}} \bigg[48 E^4 \tilde{L}^4+96 E^4 \tilde{L}^2-96 E^2 \tilde{L}^4 \nonumber \\ &\quad\quad\quad\quad -72 E^2 \tilde{L}^2+48 \tilde{L}^4-24 \tilde{L}^2+35\bigg],\\
	U_5 &= \frac{1}{256 \left(E^2-1\right)^{11/2}}\bigg[-63 -192 E^6 \tilde{L}^4+336 E^4 \tilde{L}^4  \nonumber \\ &\quad -240 E^4 \tilde{L}^2 -96 E^2 \tilde{L}^4 +200 E^2 \tilde{L}^2-48 \tilde{L}^4+40 \tilde{L}^2\bigg],
\end{align}
with $\tilde{L} := L/(2M)$. Writing
\begin{align}
	\frac{dt_p}{dr} &= \frac{dt_p}{d\tau}\left(\frac{dr_p}{d\tau}\right)^{-1},
\end{align}
we expand $dt_p/d\tau$, given in Eq.~\eqref{eq:tdot}, and combine with  \eqref{eq:urinv_exp_app} to obtain
\begin{align}         
         \frac{dt_p}{dr} &= A + \frac{2MB}{r} - \sum_{n=1}^{\infty}nC_n\left(\frac{2M}{r}\right)^{n+1},
\end{align}
and therefore 
\begin{align}
	t_p(r) = t_0 + Ar + 2MB\log\left(\frac{r}{2M}\right) + 2M\sum_{n=1}^{\infty}C_n\left(\frac{2M}{r}\right)^n.\label{eq:texp_app}
\end{align}
The dimensionless coefficients $A$ and $B$ work out as 
\begin{align}
	A &=  \frac{E}{\sqrt{E^2-1}} = \frac{1}{v},\\
	B &= \frac{E(2E^2-3)}{2(E^2-1)^{3/2}} = \frac{3v^2-1}{2v^3},
\end{align}
and the first few dimensionless coefficients $C_n$ are
\begin{align}
	C_1 &= -\frac{E  \left[4 E^2 \left(\tilde{L}^2-5\right)+8 E^4-4 \tilde{L}^2+15\right]}{8(E^2-1)^{5/2}},\\
	C_2 &= -\frac{E\left[-35 + 70 E^2 - 56 E^4 + 16 E^6 - 12 (E^2-1) \tilde{L}^2\right]}{32(E^2-1)^{7/2}},\\
	C_3 &= -\frac{E}{384(E^2-1)^{9/2}}\bigg[315 - 840 E^2 + 1008 E^4 - 576 E^6 \nonumber\\ &\quad\quad\quad + 128 E^8 + 120 (E^2-1) \tilde{L}^2 + 48 (E^2-1)^2 \tilde{L}^4\bigg],\\
	C_4 &= \frac{E}{1024(E^2-1)^{11/2}}\bigg[693 - 2310 E^2 + 3696E^4 - 3168 E^6 \nonumber\\ &\quad\quad\quad + 1408 E^8 - 256 E^{10} +  280 (-1 + E^2) \tilde{L}^2 \nonumber \\ &\quad\quad\quad + 48 (-1 + E^2)^2 (3 + 2 E^2) \tilde{L}^4\bigg],
	\\
	C_5 &= -\frac{E}{5120 \left(E^2-1\right)^{13/2}} \bigg[64 E^6 \left(5 \tilde{L}^6+15 \tilde{L}^4-429\right) \nonumber \\ &\quad\quad +1024 E^{12}-6656 E^{10}+18304 E^8 -1260 \tilde{L}^2+3003 \nonumber\\ &\quad\quad -24 E^4 \left(40 \tilde{L}^6+50 \tilde{L}^4-1001\right) +720 \tilde{L}^4 \nonumber \\ &\quad\quad+12 E^2 \left(80 \tilde{L}^6-40 \tilde{L}^4+105 \tilde{L}^2-1001\right) -320 \tilde{L}^6 \bigg].
\end{align}
In Eq.~\eqref{eq:texp_app} $t_0$ is a constant of integration, whose value is fixed by the initial condition imposed on $t_p$. In practice we determine $t_0$ by comparing the expansion \eqref{eq:texp_app} to a numerical integration of Eq.~\eqref{eq:dt_dchi} at large radii. 

Similarly, we can express $\vf_p(r)$ along the outbound leg of the orbit as an expansion in $1/r$ at large $r$. The result is
\begin{align}
	\varphi_p(r) &= \varphi_{out} + \displaystyle\sum_{n=1}^{\infty}D_n\left(\frac{2M}{r}\right)^n,
\end{align}
where $\vf_{out} := \varphi_p(\chi_{\infty})$ and the first few coefficients are
\begin{align}
	D_1 &= -\frac{\tilde{L}}{\sqrt{E^2-1}},\\
	D_2 &= \frac{\tilde{L}}{4 \left(E^2-1\right)^{3/2}},\\
	D_3 &= \frac{\tilde{L} \left(-4 E^2 \tilde{L}^2+4 \tilde{L}^2-3\right)}{24 \left(E^2-1\right)^{5/2}},\\
	D_4 &= \frac{\tilde{L} \left[8 E^2 \left(E^2-1\right) \tilde{L}^2+4 \left(E^2-1\right) \tilde{L}^2+5\right]}{64 \left(E^2-1\right)^{7/2}},\\
	D_5 &= -\frac{\tilde{L}}{1920 \left(E^2-1\right)^{9/2}}\bigg[144 \left(E^2-1\right)^2 \tilde{L}^4 \nonumber\\ &\quad\quad+288 E^2 \left(E^2-1\right) \tilde{L}^2+72 \left(E^2-1\right)\tilde{L}^2+105\bigg].
\end{align}

\section{Boundary conditions for homogeneous solutions}\label{app:homsol_BCs}
To obtain boundary conditions for the homogeneous solution $\psi_{\lw}^+(r)$ in the limit $r \rightarrow\infty$, we make the ansatz \cite{Barack:2008ms}
\begin{align}
	\psi_{\lw}^+(x) &= e^{i\omega r_*}\sum_{k = 0}^{k_{\rm out}} c_k^{\infty}\left(\frac{2M}{r}\right)^{k},
\end{align}
and substitute into the homogeneous equation \eqref{eq:ScalarEOMFD_hom}. This yields the recurrence relation 
\begin{align}
	\sum_{i = 0}^3 f_i^{\infty} c_{k-i}^{\infty} = 0 \label{eq:recurr_infty},
\end{align}
where the coefficients are given by
\begin{align}
	f_0^{\infty} &= -2i\tilde{\omega}k,\\
	f_1^{\infty} &= k(k-1) +2i\tilde{\omega}(k-1)-l(l+1),\\
	f_2^{\infty} &= -2k^2 + 5k - 3 +l(l+1),\\
	f_3^{\infty} &= (k-2)^2,
\end{align}
with
$\tilde{\omega} := 2M\omega$. 

Near the horizon we have a similar expansion for $\psi_{\lw}^-(r)$,
\begin{align}
	\psi_{\lw}^-(x) &= e^{-i\omega r_*}\sum_{k = 0}^{k_{\rm in}} c_k^{\text{eh}}y^k, 
\end{align}
where $y := r/2M - 1$. Again substituting into Eq.~\eqref{eq:ScalarEOMFD_hom}, we arrive at a recurrence relation for the coefficients,
\begin{align}
	\sum_{i = 0}^3 d_i^{\rm eh} c_{k-i}^{\rm eh} = 0, \label{eq:recurr_hor}
\end{align}
with
\begin{align}
	d_0^{\rm eh} &= -2i\tilde{\omega}k + k^2,\\
	d_1^{\rm eh} &= -6i\tilde{\omega}(k-1)-1-l(l+1)+(k-1)(2k-3),\\
	d_2^{\rm eh} &= (k-2)(k-3) -l(l+1) -6i\tilde{\omega}(k-2),\\
	d_3^{\rm eh} &= -2i\tilde{\omega}(k-3).	
\end{align}

The recurrence relations \eqref{eq:recurr_infty} and \eqref{eq:recurr_hor} admit solutions with $c_{k<0}^{\infty/{\rm eh}} = 0$, $c_0^{\infty/{\rm eh}} = 1$, and $c_k^{\infty/{\rm eh}}$ determined recursively. The number of terms, and hence accuracy of the approximation, may be controlled by varying the truncation parameters $k_{\rm out}$ and $k_{\rm in}$.

\section{Constants appearing in the tail correction scheme}\label{app:correction_scheme_constants}
In this appendix we collect the various coefficients that appear in the expansions that make up the tail correction scheme described in Section~\ref{sec:correction_scheme}. 
The first 5 coefficients $\hat{c}_n^{\infty}$ appearing in Eq.~\eqref{c_hat} are given 
in terms of the coefficients $c_n^{\infty}$ of Eq.~\eqref{eq:asymp_inf} by
\begin{align}
	\hat{c}_1^{\infty} &:= c_1^{\infty} - i\tilde{\omega},\\
	\hat{c}_2^{\infty} &:= c_2^{\infty} - i\tilde{\omega}c_1^{\infty} + \frac{1}{2}(-i\tilde{\omega} - \tilde{\omega}^2),\\
	\hat{c}_3^{\infty} &:= \frac{1}{6}\big[-3 c_1^{\infty} \tilde{\omega}^2-3 i c_1^{\infty} \tilde{\omega}-6 i c_2^{\infty} \tilde{\omega}+6 c_3^{\infty}+i \tilde{\omega}^3 \nonumber\\ &\quad\quad\quad\quad -3 \tilde{\omega}^2-2 i \tilde{\omega}\big],\\
	\hat{c}_4 ^{\infty} &:= \frac{1}{24}\bigg[4 i \tilde{\omega}^3 c_1^{\infty }-12 \tilde{\omega}^2 c_1^{\infty }-12 \tilde{\omega}^2 c_2^{\infty }-8 i \tilde{\omega} c_1^{\infty }-12 i \tilde{\omega} c_2^{\infty }\nonumber\\ &\quad\quad\quad-24 i \tilde{\omega} c_3^{\infty }+\tilde{\omega}^4+6 i \tilde{\omega}^3-11 \tilde{\omega}^2-6 i \tilde{\omega}+24 c_4^{\infty }\bigg],\\
	\hat{c}_5^{\infty} &:= \frac{1}{120}\bigg[5 \tilde{\omega}^4 c_1^{\infty }+30 i \tilde{\omega}^3 c_1^{\infty }+20 i \tilde{\omega}^3 c_2^{\infty }-55 \tilde{\omega}^2 c_1^{\infty }\nonumber \\&\quad\quad-60 \tilde{\omega}^2 c_2^{\infty }-60 \tilde{\omega}^2 c_3^{\infty }-30 i \tilde{\omega} c_1^{\infty } -40 i \tilde{\omega} c_2^{\infty }\nonumber\\ &\quad\quad-60 i \tilde{\omega} c_3^{\infty }-120 i \tilde{\omega} c_4^{\infty }-i \tilde{\omega}^5+10 \tilde{\omega}^4+35 i \tilde{\omega}^3\nonumber\\ &\quad\quad\quad-50 \tilde{\omega}^2-24 i \tilde{\omega}+120 c_5^{\infty }\bigg].
\end{align}
The first 5 coefficients $H_{n\sigma}$ appearing in Eq.~\eqref{eq:corr_exponential_expansion} are given
in terms of the quantities $\Delta_n$ of Eq.~\eqref{Delta_inf} by
\begin{align}
	H_{1,\sigma} &:= i \sigma  \Delta_{\infty }^{(1)},\\
	H_{2,\sigma} &:= \frac{1}{2} \left[-\left(\Delta_1\right)^2+2 i \sigma  \Delta_2\right]\\
	H_{3,\sigma} &:= \frac{i}{6} \left[-\sigma  \left(\Delta_1\right)^3+6 \sigma  \Delta_3+6 i \Delta_2 \Delta_1\right],\\
	H_{4,\sigma} &:= \frac{1}{24} \bigg[-12 i \sigma  \Delta_2 \left(\Delta_1\right)^2+24 i \sigma  \Delta_4+\left(\Delta_1\right)^4\nonumber\\ &\quad\quad\quad-24 \Delta_3 \Delta_1-12 \left(\Delta_2\right)^2\bigg],\\
	H_{5,\sigma} &:= \frac{i}{120}  \bigg[\sigma \left(\Delta_1\right)^5-20 i \Delta_2 \left(\Delta_1\right)^3\nonumber\\ &\quad\quad-60 \sigma \Delta_3 \left(\Delta_1\right)^2 -60 \sigma \left(\Delta_2\right)^2 \Delta_1\nonumber\\ &\quad\quad + 120 i \Delta_4 \Delta_1+120 i\Delta_2 \Delta_3+120 \sigma  \Delta_5\bigg].
\end{align}

~\\~\\

\bibliographystyle{unsrt}
\bibliography{biblio}

\end{document}